\documentclass[final,authoryear,5p,times,twocolumn]{elsarticle}

\pdfoutput=1

\usepackage{graphicx}
\usepackage[labelformat=empty]{subfig}

\usepackage{amssymb}

\usepackage[pdftex,pdfpagemode={UseOutlines},bookmarks,bookmarksopen,colorlinks,linkcolor={blue},citecolor={green},urlcolor={red}]{hyperref}
\usepackage{hypernat}

\journal{Astronomy \& Computing}

\usepackage{upquote}

\begin{document}

\begin{frontmatter}



\title{FellWalker - a Clump Identification Algorithm }


\author[jac]{David S.\ Berry\corref{cor1}}
\ead{d.berry@jach.hawaii.edu}

\cortext[cor1]{Corresponding author}

\address[jac]{Joint Astronomy Centre, 660 N.\ A`oh\=ok\=u Place, Hilo, HI
  96720, USA}

\begin{abstract}
This paper describes the FellWalker algorithm, a \emph{watershed}
algorithm that segments a 1-, 2- or 3-dimensional array of data values
into a set of disjoint clumps of emission, each containing a single
significant peak. Pixels below a nominated constant data level are
assumed to be background pixels and are not assigned to any clump.
FellWalker is thus equivalent in purpose to the CLUMPFIND algorithm.
However, unlike CLUMPFIND, which segments the array on the basis of a set
of evenly-spaced contours and thus uses only a small fraction of the
available data values, the FellWalker algorithm is based on a
gradient-tracing scheme which uses all available data values. Comparisons
of CLUMPFIND and FellWalker using a crowded field of artificial Gaussian
clumps, all of equal peak value and width, suggest that the results
produced by FellWalker are less dependent on specific parameter settings
than are those of CLUMPFIND.
\end{abstract}


\end{frontmatter}


\newcommand{\mnras}{Mon Not R Astron Soc}
\newcommand{\aap}{Astron Astrophys}
\newcommand{\aaps}{Astron Astrophys Supp}
\newcommand{\pasp}{Pub Astron Soc Pacific}
\newcommand{\apj}{Astrophys J}
\newcommand{\apjs}{Astrophys J Supp}
\newcommand{\qjras}{Quart J R Astron Soc}
\newcommand{\an}{Astron.\ Nach.}
\newcommand{\ijimw}{Int.\ J.\ Infrared \& Millimeter Waves}
\newcommand{\procspie}{Proc.\ SPIE}
\newcommand{\aspconf}{ASP Conf. Ser.}

\newcommand{\ascl}[1]{\href{http://www.ascl.net/#1}{ascl:#1}}

\section{Introduction}
\label{sec:intro}

The CLUMPFIND algorithm \citep[][\ascl{1107.014}]{1994Williams} has been
widely used for decomposing 2- and 3-dimensional data into disjoint
clumps of emission, each associated with a single significant peak. It is
based upon an analysis of a set of evenly spaced contours derived from
the data array and has two main parameters - the lowest contour level,
below which data is ignored, and the interval between contours. However
it has often been noted (\emph{e.g.} \citet{2014Christie},
\citet{2009Kainulainen}, \citet{2009Pineda}, \citet{2008Smith}, \citet{2007Elia},
\citet{2003Brunt}) that the decomposition produced by CLUMPFIND can be very
sensitive to the specific value used for the contour interval, particularly
for 3-dimensional data and crowded fields. The choice of an optimal contour
interval is a compromise - real peaks may be missed if the interval is too
large, but noise spikes may be interpreted as real peaks if the interval is
too small.

The FellWalker algorithm attempts to circumvent these issues by avoiding
the use of contours altogether. Only a small fraction of the available
pixel values fall on the contour levels used by CLUMPFIND - the majority
fall \emph{between} these levels and so will have no effect on the
resulting decomposition. By contrast, FellWalker makes equal use of all
available pixel values above a stated threshold.

\begin{figure}
\includegraphics[width=\columnwidth]{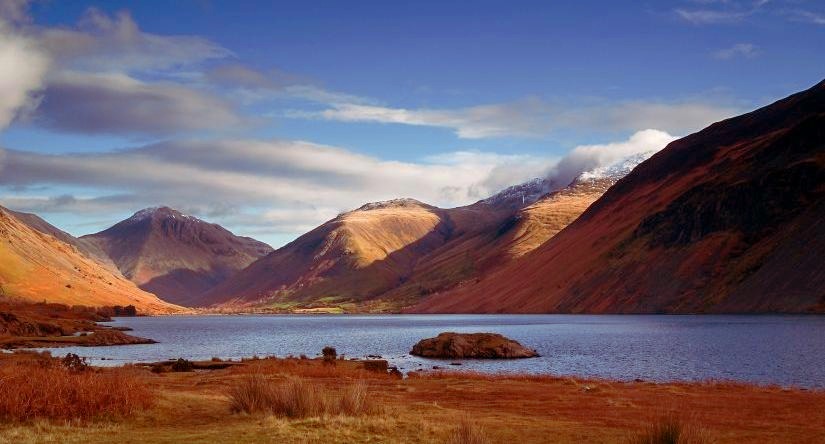}
\caption{Wastwater and the Wasdale Fells, including Great Gable
(centre-left) and snow-covered Scafell Pike, the highest point in England
at 978 metres above sea level, just visible under cloud. Copyright: Nick
Thorne, \url{http://www.lakedistrict.gov.uk/learning/freephotos\#} }
\label{fig:wasdale}
\end{figure}

The name ``Fell Walker'' relates to the popular British pass-time of
walking up the hills and mountains of northern England, particularly
those of the Lake District (see Fig.~\ref{fig:wasdale} and
\htmladdnormallink{http://en.wikipedia.org/wiki/Hillwalking}~), and was
chosen to reflect the way in which the algorithm proceeds iteratively by
following an upward path from a low-valued pixel to a significant summit
or peak in data-value. The following description of the algorithm uses
this fell-walking metaphor at frequent intervals.

FellWalker is a form of \emph{watershed} algorithm \citep{2001Roerdink} -
a class of algorithms that segment images by identifying the
``watershed'' lines that separate low lying areas (``catchment basins'').
FellWalker inverts this idea so that each identified segment of the array
is associated with a peak (a local maximum), rather than a basin (a local
minimum). It shares much in common with the HOP algorithm
\citep{1998Eisenstein} - another gradient-tracing watershed algorithm, but is
designed for use with gridded observational data rather than particles in
an N-body simulation. HOP is known to be relatively insensitive to the
values supplied for its parameters (except the lower threshold),
reflecting a similar feature found for FellWalker (see
section~\ref{sec:compare}).

An implementation of the FellWalker algorithm is included in the Starlink
\htmladdnormallinkfoot{CUPID}{http://starlink.jach.hawaii.edu/starlink/CUPID}
package \citep[][\ascl{1311.007}]{CupidAdass,SUN255}, together with
implementations of other clump-finding algorithms such as GaussClumps
\citep[][\ascl{1406.018}]{1990ApJ...356..513S} and CLUMPFIND. In common
with the rest of the Starlink software
\citep[][\ascl{1110.012}]{StarlinkAdass}, the source code for the CUPID
package is open-source and is available on
\htmladdnormallinkfoot{Github}{https://github.com/Starlink}. Pre-built
binaries for the complete Starlink software collection can be obtained
from the \htmladdnormallinkfoot{Joint Astronomy Centre, Hawaii}
{http://starlink.jach.hawaii.edu/starlink}. The native data format used
by CUPID is the Starlink NDF \citep{2014Jenness}, but FITS data can be handled
transparently by means of the Starlink CONVERT package \citep{SUN55}.

\begin{figure}
\includegraphics[width=\columnwidth]{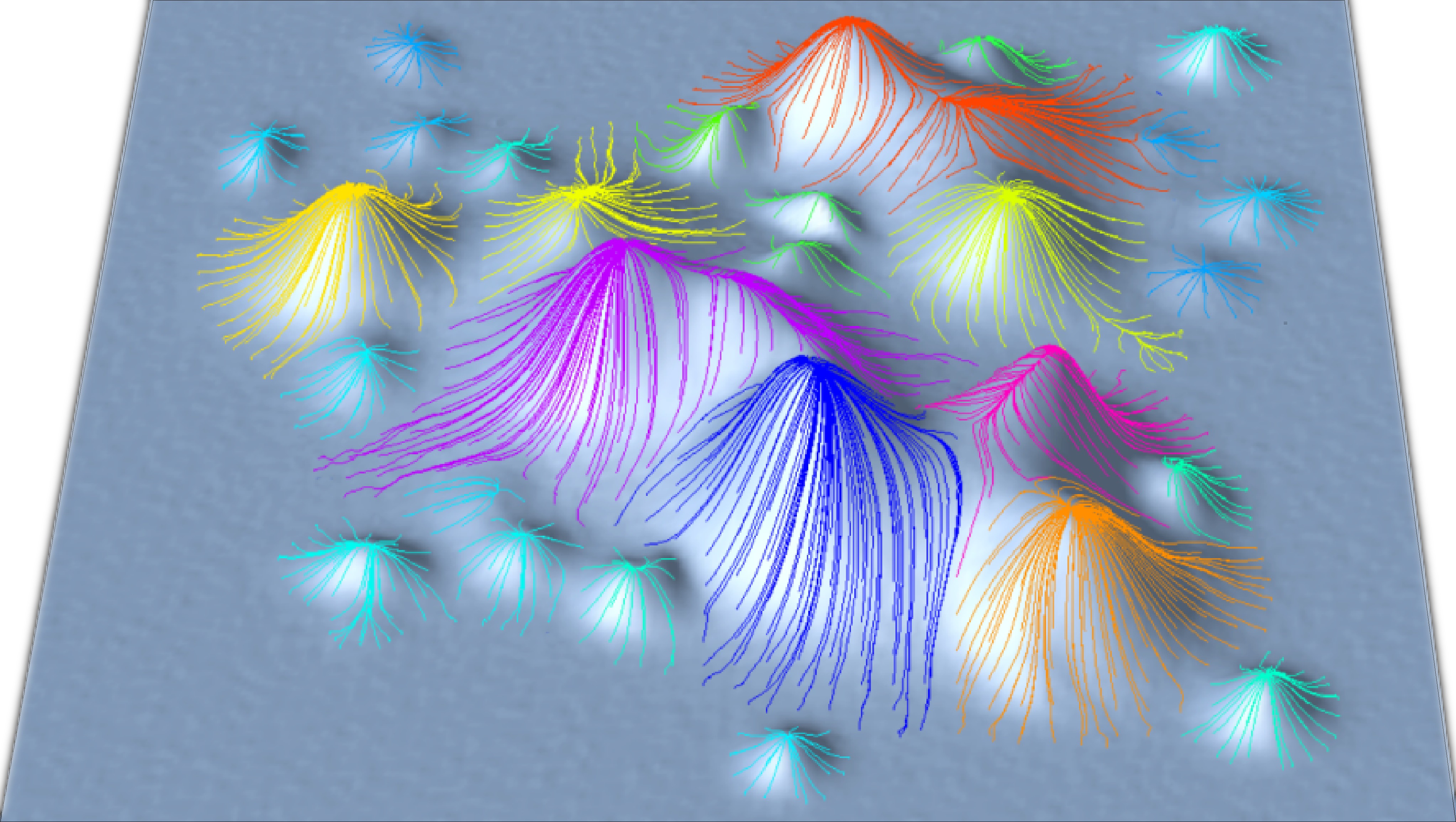}
\caption{In 2-dimensions, peaks in data value are often reminiscent of the
fells of northern England such as those in Fig.~\ref{fig:wasdale}. The
FellWalker algorithm performs many walks starting at various low-land
pixels, and for each one follows a line of steepest ascent until a
significant summit is reached. All walks that terminate at the same peak
are assigned to the same clump, indicated by different colours in the
above figure.}
\label{fig:fellwalking}
\end{figure}

FellWalker, like CLUMPFIND, segments the supplied data array into a
number of disjoint regions, each associated with a single significant
peak. Whilst this approach has been used widely, there are several
alternative approaches to the problem of identifying clumps of emission
which can be more appropriate, depending on the particular science being
performed. For instance, it may be beneficial to allow clumps of emission
to overlap (\emph{e.g.} GaussClumps \citep[][\ascl{1406.018}]{1990ApJ...356..513S}
and GetSources \citep{GetSources}), or to take account of the hierarchical
structure within clouds (\emph{e.g.} dendrograms \citep{2008Rosolowsky}). However,
such issues are outside the remit of the FellWalker algorithm, and consequently
this paper provides only a comparison of FellWalker with CLUMPFIND.

\section{The FellWalker Algorithm}

The core of the FellWalker algorithm consists of following many different
paths of steepest ascent in order to reach a significant summit, each of
which is associated with a clump, as illustrated in Fig.~\ref{fig:fellwalking}.
Every pixel with a data value above a user-specified threshold is used in
turn as the start of a ``walk''. A walk consists of a series of steps,
each of which takes the algorithm from the current pixel to an immediately
neighbouring pixel of higher value, until a pixel is found which is
higher than any of its immediate neighbours. When this happens, a search
for a higher pixel is made over a larger neighbourhood. If such a pixel
is found the walk jumps the gap and continues from this higher pixel. If no
higher pixel is found it is assumed that a new summit has been reached ---
a new clump identifier is issued and all pixels visited on the walk are
assigned to the new clump. If at any point a walk encounters a pixel
which has already been assigned to a clump, then all pixels so far
visited on the walk are assigned to that same clump and the walk
terminates.

It is possible for this basic algorithm to fragment up-land plateau
regions into lots of small clumps which are well separated spatially but
have only minimal dips between them. The raw clumps identified by the above
process can be merged to avoid such fragmentation, on the basis of a
user-specified minimum dip between clumps\footnote{In common with other
parameters, this minimum dip parameter is specified as a multiple of the
noise level in the data.}. These merged clumps may,
optionally, be cleaned by smoothing their boundaries using a single step
of a cellular automaton.

Finally, each clump is characterised using a number of statistics, and a
catalogue of clumps statistics is created together with a pixel mask
identifying the clump to which each pixel is assigned.

The following sections give more detailed descriptions of each of these
phases in the FellWalker algorithm.

\subsection{Identifying Raw Clumps}
\label{sec:raw}
An array of integer values is first allocated, which is the same shape
and size as the supplied data array. This ``clump assignment array''
(CAA) is used to record the integer identifier of the clump, if any, to
which each pixel has been assigned. All clump identifiers are greater
than zero. An initial pass is made through the supplied data array to
identify pixels which have a data value above a user-specified threshold
value. Such pixels are assigned a value of zero in the CAA indicating
that the pixel is usable but has not yet been assigned to a clump, and
all other pixels are assigned a value of $-1$ indicating that they are
unusable and should never be assigned to a clump.

This initial CAA is then searched for any isolated individual pixels
above the threshold. Such pixels are set to $-1$ in the CAA, indicating
they should be ignored.

The main loop is then entered, which considers each pixel in turn as the
potential start of a walk to a peak. Pixels which have a non-zero value
in the CAA are skipped since they have either already been assigned to a
clump (if the CAA value is positive) or have been flagged as unusable (if
the CAA value is negative). A single walk consists of stepping from pixel
to pixel until a pixel is reached which is already known to be part of a
clump, or a significant isolated peak is encountered. The vector indices of
the pixels visited along a walk are recorded in a temporary array so that
they  can be identified later.

At each step, the pixel values within a box of width three pixels are compared
to the central pixel to find the neighbouring pixel which give the highest
gradient\footnote{This gradient takes into account the fact that the
centres of the corner pixels are further away from the box centre than
are the centres of the mid-side pixels.}. Thus 2 neighbours are checked if
the data is 1-dimensional data, 8 are checked if the data is 2-dimensional
and 26 are checked if the data is 3-dimensional. The gradient is
evaluated in pixel coordinates, without regard to the physical units
associated with each axis.

If the highest gradient found above is greater than zero --- that is, if
there is an upward route out of the current pixel --- the walk steps to the
selected neighbouring pixel. If this new pixel has already been assigned
to a clump (\emph{i.e.} if the CAA holds a positive value at the new
pixel), then the new walk has joined an older walk and so will eventually
end up at the same peak as the older walk. In this case, the existing
positive CAA value of the new pixel (\emph{i.e.}, the clump index assigned
to the older walk) is copied into the CAA for all pixels visited so far
on the new walk, and a new walk from the next starting pixel is
initiated.

If the highest gradient found to any neighbouring pixel is less than or
equal to zero, then there is no upward route from the central pixel. This
could mean the walk has reached a significant peak, but it could also
mean it has merely reached a noise spike. To distinguish these two cases,
a search is made over a larger box\footnote{By default a box of width 9
pixels, but the user can specify a different size.}. If the maximum pixel
value in this larger box is smaller than the central pixel value, then
the central pixel is considered to be a significant peak. A new clump
identifier is issued for it and stored in the CAA at all pixels visited
on the walk. A new walk from the next starting pixel is then initiated.

If the maximum pixel value found in the larger box is greater than the
central pixel value, then the central pixel is considered to be a noise
spike. The walk then ``jumps across the gap'' and continues from the
highest pixel found in the box.

\begin{figure}
\includegraphics[width=\columnwidth]{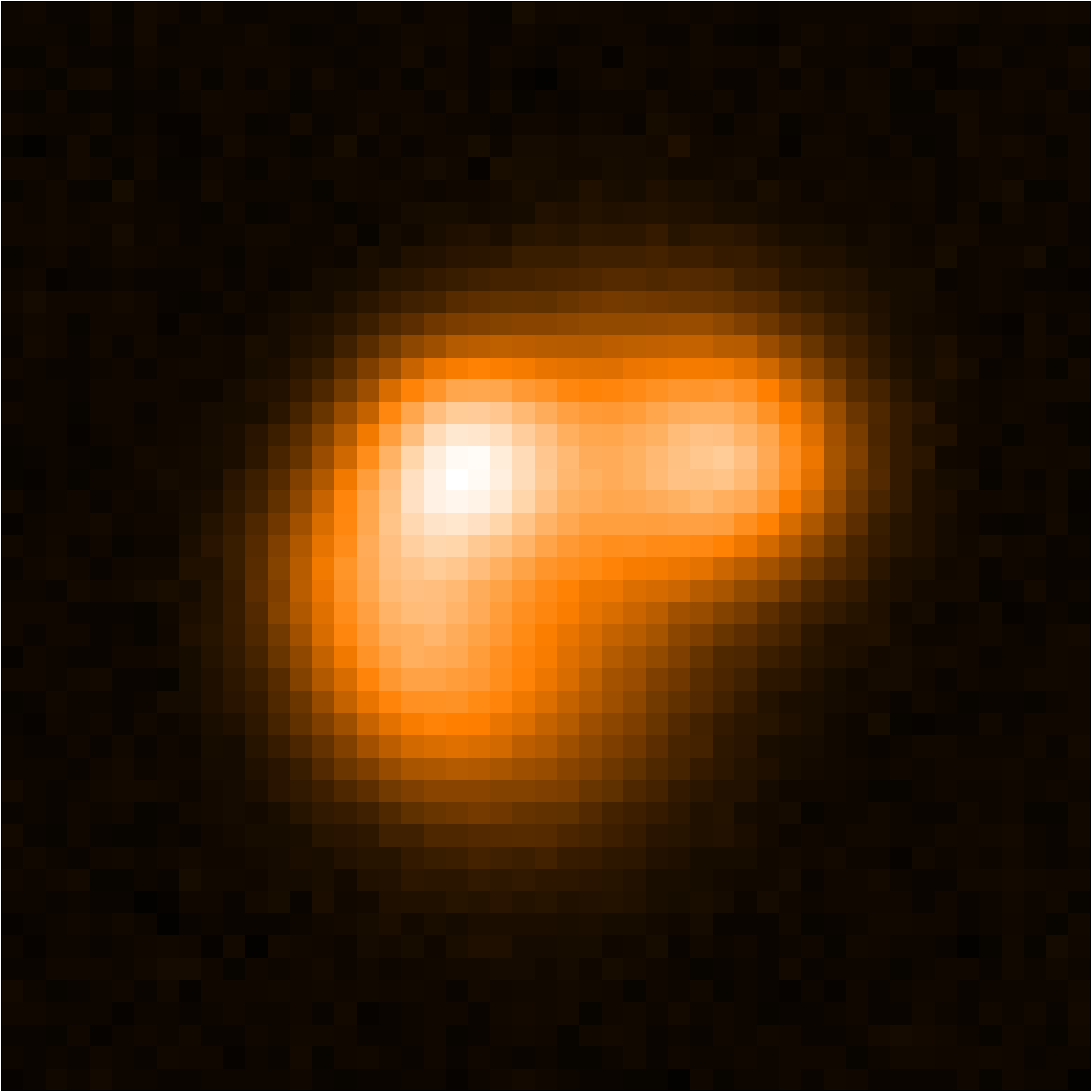}
\caption{A $50\times50$ array of artificial data used to illustrate the
FellWalker algorithm below.}
\label{fig:sim}
\end{figure}

\begin{figure}
\includegraphics[width=\columnwidth]{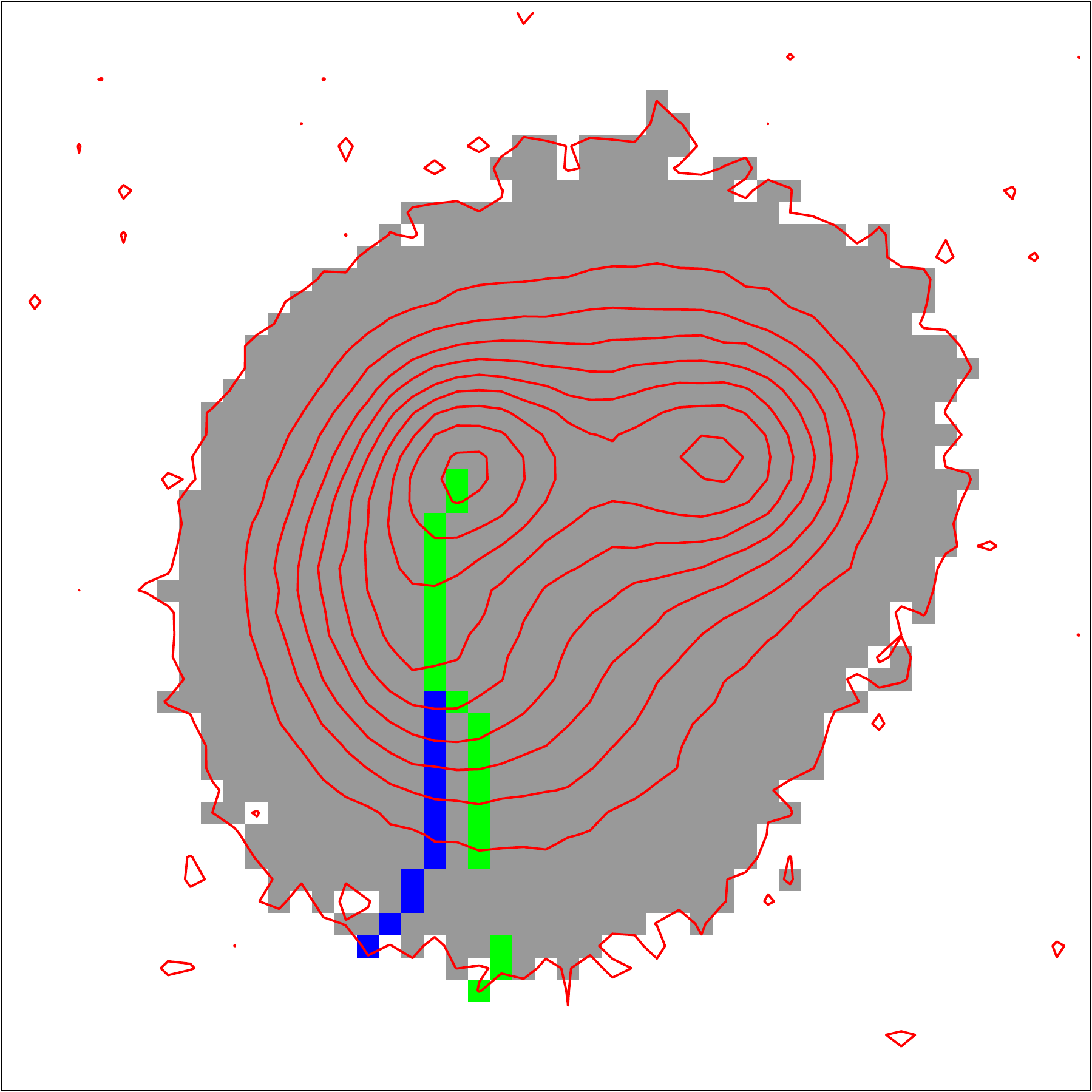}
\caption{Two walks to a peak within the artificial data shown in
Fig.~\ref{fig:sim}. The contours show the data values themselves. The white
background pixels are below the nominated threshold, the grey pixels are
above the threshold but have not yet been assigned to a clump. The green
pixels trace the first walk that reached the left-hand peak. The blue pixels
trace a later walk to the same peak that was terminated when it met the first
walk. The green and blue pixels are all assigned to the same clump. These
walks follow the steepest line of ascent. Note the gap in the green line
near its start at the lowest contour --- this is where a jump was made from
a noise spike to the highest value in a $9\times9$ box of neighbouring pixels.}

\label{fig:walks}
\end{figure}

The above process results in the CAA holding a clump identifier for every
usable pixel in the supplied data array. However, some of the walks
performed above may start with a section of very low gradient before any
significant ascent begins. The user is allowed to specify a minimum
gradient which must be achieved before a walk is considered to have
begun. Any section of the walk that occurs before the first such
``steep'' section is flagged as unusable in the CAA. For this test,
the gradient of a walk is averaged over four consecutive steps.

\begin{figure}
\includegraphics[width=\columnwidth]{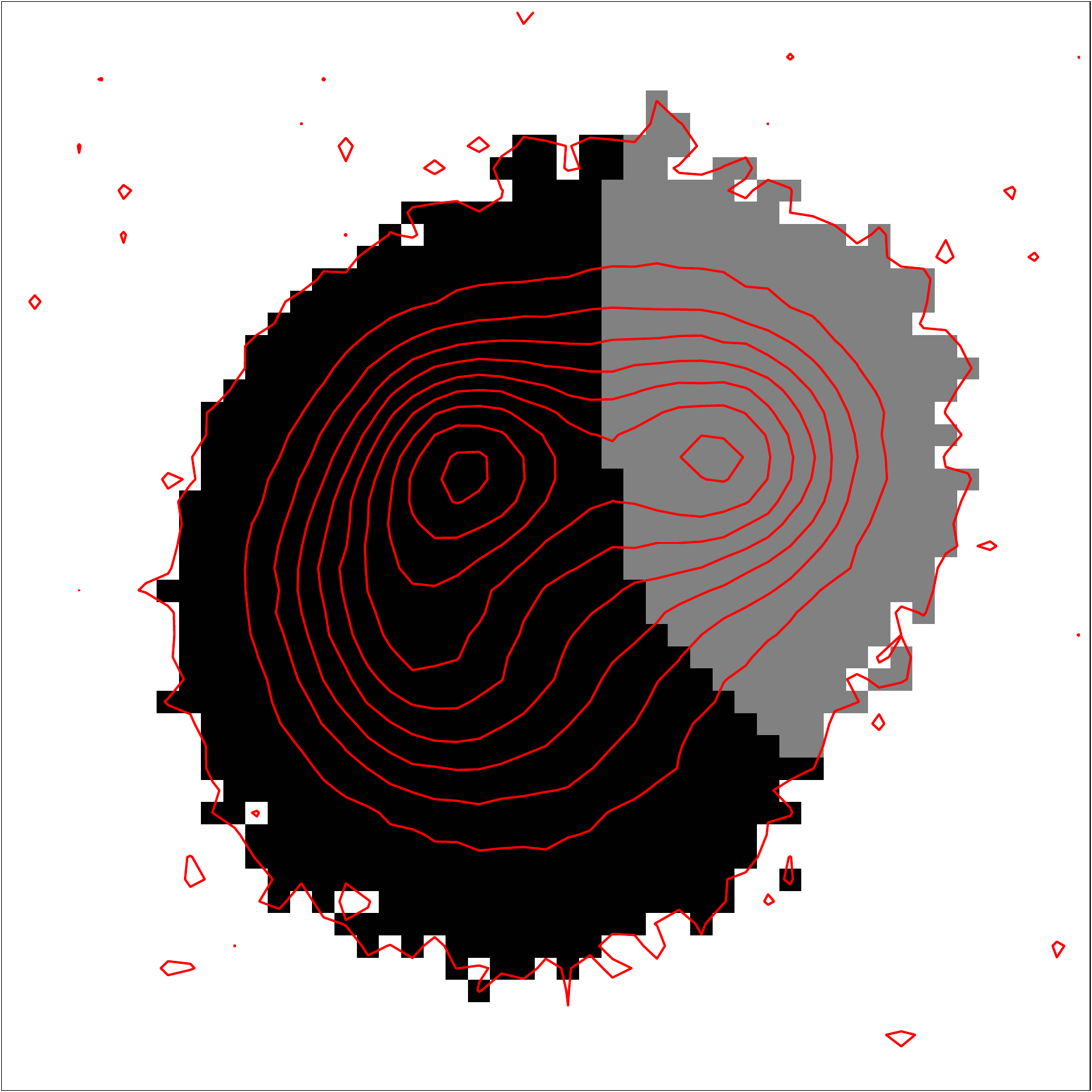}
\caption{The raw clump mask created for the artificial data shown in
Fig.~\ref{fig:sim}. Two clumps are found, indicated by the black and grey
pixels. Walks starting from the black pixels all terminate at the left
hand peak, and walks starting from the grey pixels all terminate at the
right hand peak.}
\label{fig:rawmask}
\end{figure}

This process is illustrated in Fig.~\ref{fig:walks} which shows two
example up-hill walks produced by FellWalker for the artificial data shown
in Fig.~\ref{fig:sim}. The final CAA produced by the above process for
this data is shown in Fig.~\ref{fig:rawmask}.

\subsection{Merging Clumps}

The number of significant peaks found by the above process is determined
primarily by the maximum distance a walk can jump when searching for a
higher neighbouring pixel value. This parameter --- known as \emph{MaxJump}
--- defaults to 4 pixels. Using a larger value results in more local peaks
being interpreted as noise spikes rather than significant peaks, with a
corresponding reduction in the number of significant peaks found. Thus at
this point, peaks are discriminated simply on the basis of their spatial
separation.

This means it is possible for a clump with a wide, flat summit to be
fragmented into multiple clumps on the basis of noise spikes that are
separated by more than \emph{MaxJump} pixels.

To correct this, FellWalker merges adjacent clumps if the ``valley''
between the two adjacent peaks is very shallow.

Each clump (referred to as the ``central'' clump below) is considered in
turn to see if it should be merged with any of its neighbouring clumps.
The height of the ``col\footnote{The highest point on the boundary
between the two clumps.}'' between the central clump and each
neighbouring clump is found in turn, and the neighbouring clump with the
highest col is selected as a candidate for merging. If the peak value in
the central clump is less than a specified value,
\emph{MinDip}\footnote{The default is three times the noise level in the
data.}, above the col, the two clumps are merged into a single clump.

Once all central clumps have been checked in this way, the whole process
is repeated to see if any of the merged clumps should themselves be
merged. This process repeats until no further clumps can be merged.

\subsection{Cleaning Clump Outlines}
\label{sec:cleaning}

Once neighbouring clumps separated by shallow valleys have been merged,
there is an option to smooth the boundaries between adjacent clumps to
reduce the effects of noise. This is done using a specified number of
steps of a cellular automaton to modify the integer values in the CAA.

\begin{figure}
\includegraphics[width=\columnwidth]{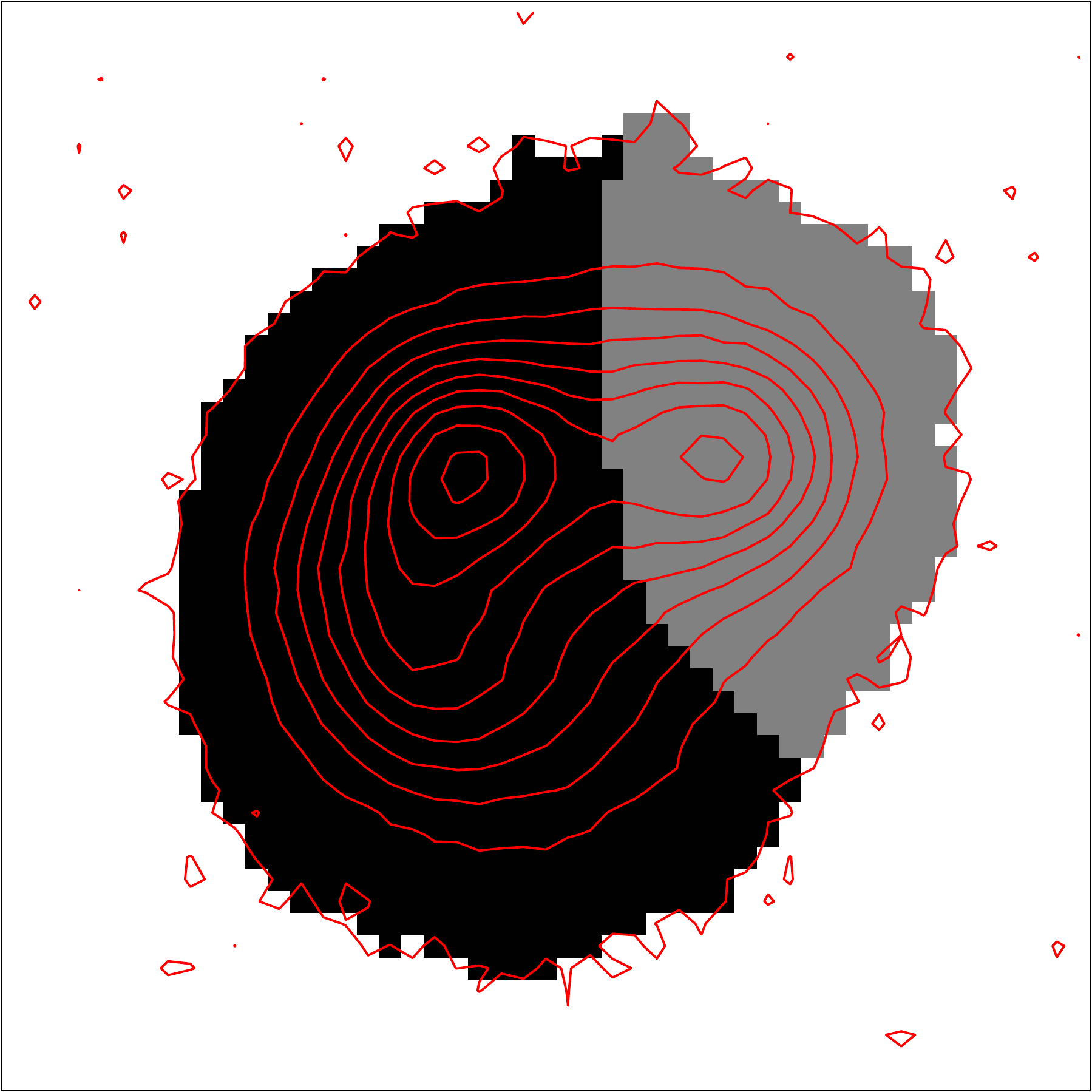}
\caption{The smoothing effect of a single step of the cellular automaton
on the clump outlines shown in Fig.~\ref{fig:rawmask}. The outer
boundaries of the two clumps are noticably smoother, although in this
particular case the boundary \emph{between} the two clumps has not
changed much, since it is already very smooth.}
\label{fig:cleanedmask}
\end{figure}

A single step of the cellular automaton creates a new CAA from the old
CAA. Each pixel in the new CAA is set to the most commonly occurring
clump index within a box of width 3 pixels centred on the corresponding
pixel within the old CAA. The output CAA from one step becomes the input
CAA to the next step. By default, only one step is performed.
Fig.~\ref{fig:cleanedmask} shows the effects of applying a single step to
the CAA shown in Fig.~\ref{fig:rawmask}. Note, this cleaning process
produces only minimal changes in the smooth artificial data used in these
figures. The effect of the cleaning process can be much more pronounced
in real data.

\subsection{Removing Unusable Clumps}
Various criteria are available to select unusable clumps and exclude them
from the final data products. These include:

\begin{itemize}
\item Clumps that touch an edge of the supplied data array.
\item Clumps that touch areas of missing (\emph{i.e.} blank) pixels.
\item Clumps that have a peak value less than a given limit.
\item Clumps that contain fewer than a given number of pixels.
\end{itemize}

The number of clumps rejected for each of these reasons is reported.

\subsection{Characterising Each Clump}
The FellWalker algorithm is implemented within the {\tt findclumps}
command of the Starlink CUPID package. This
command implements several other clump finding algorithms in addition to
FellWalker, and one of its design requirements was that each algorithm should characterise clumps in
the same way, so that results from different algorithms can be compared
directly. The results of each algorithm are presented in the following ways:

\begin{enumerate}

\item A pixel mask which is the same shape and size as the supplied data
array. Each pixel value is an integer which gives the index of the clump
to which the pixel has been assigned. Pixels that have not been assigned
to any clump are flagged with a special value.

\item A set of minimal cut-outs from the supplied data array. Each cut-out
holds the supplied pixels values corresponding to a single clump, with pixels
outside the clump set to a special ``blank'' value.

\item A table in which each row describes a single clump. The columns are:

\begin{description}
\item[Peak1] The position of the clump peak value on axis 1.
\item[Peak2] The position of the clump peak value on axis 2.
\item[Peak3] The position of the clump peak value on axis 3.
\item[Cen1] The position of the clump centroid on axis 1.
\item[Cen2] The position of the clump centroid on axis 2.
\item[Cen3] The position of the clump centroid on axis 3.
\item[Size1] The size of the clump along pixel axis 1.
\item[Size2] The size of the clump along pixel axis 2.
\item[Size3] The size of the clump along pixel axis 3.
\item[Sum] The total data sum in the clump (\emph{i.e.} the sum of the
pixel values within the clump).
\item[Peak] The peak value in the clump.
\item[Volume] The total number of pixels falling within the clump.
\item[Shape] An optional column containing an STC-S description
\citep{2007STCS,2007STC,2010ASPC..434..213B} of the spatial coverage
of the clump. STC-S is standard developed by the International Virtual
Observatory Alliance\footnote{http://www.ivoa.net} to describe Space-Time
Coordinate (STC) metadata in the form of a simple linear string of
characters. In the case of CUPID, this metadata describes either a
polygonal or elliptical region within the spatial World Coordinate
System (WCS) of the supplied input data.
\end{description}

\end{enumerate}

The values stored in the size columns of the output table are the RMS
deviation of each pixel centre from the clump centroid, where each pixel
is weighted by the corresponding pixel data value minus an estimate of
the background value in the clump\footnote{The minimum data value in the
clump is used as the background value.}. So for each axis, the size of
the clump on that axis is given by:

\[ size = \sqrt{ \frac{ \sum d_{i}.x_{i}^{2} }{ \sum d_{i} } -
\left( \frac{\sum d_{i}.x_{i} }{\sum d_{i}}  \right)^2 } \]

where $d_{i}$ is the data value of pixel $i$ minus the background value,
and $x_{i}$ is the axis value of pixel $i$. For a clump with a Gaussian
profile, this ``size'' value is equal to the standard deviation of the
Gaussian.

For observational data, the clumps will be blurred by the telescope beam.
FellWalker includes an option to remove this blurring if the beam size of
the telescope is known:

\[ size_{corrected} = \sqrt{ size^{2} - beam^{2} } \]

where $beam$ is the standard deviation of the Gaussian beam profile. A
corresponding correction is also applied to the peak values stored within
the table, on the assumption that the product of the peak value and the
clump volume should be unchanged by the instrumental blurring:

\[ peak_{corrected} = peak.(size/size_{corrected}) \]

\section{Comparing FellWalker and CLUMPFIND}
\label{sec:compare}
\citet{2010Watson} made a detailed comparison of the performance of several
different 2-dimensional clump finding algorithms, including FellWalker and CLUMPFIND.
This concluded that, under the conditions used in the study, FellWalker
is less likely than CLUMPFIND to split up large clumps, and is less
likely to create false clump detections. A similar conclusion was reached
by \citet{2014Christie} for 3-dimensional data. The tendency for CLUMPFIND
to split sources has also been noted by \citet{2006Enoch}.

\begin{figure}
\includegraphics[width=\columnwidth]{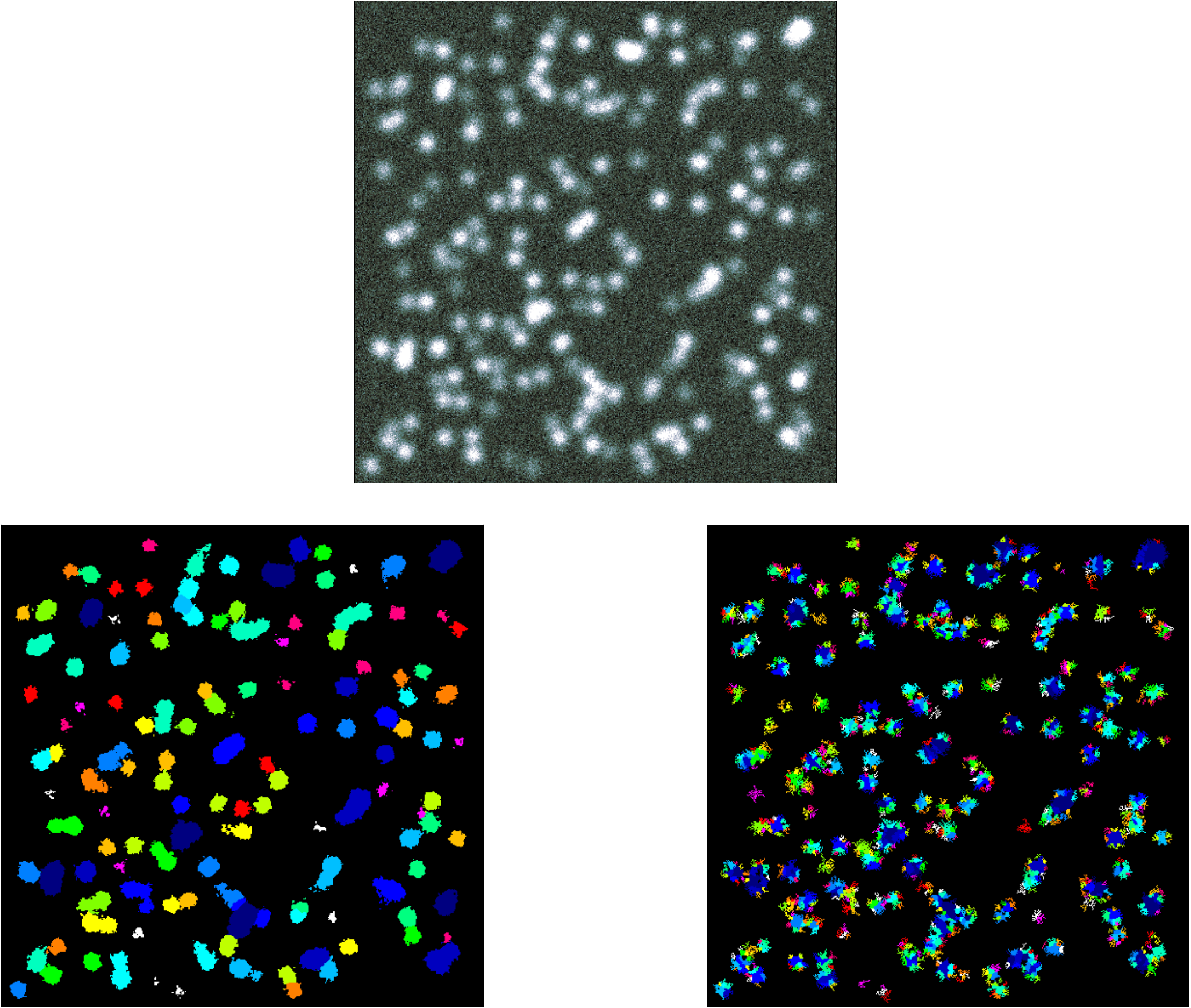}
\caption{Top: field of artificial Gaussian clumps. Lower left: the clump
assignment array produced by FellWalker. Lower right: the clump assignment
array produced by CLUMPFIND. Each colour indicates a different clump. }
\label{fig:comp1}
\end{figure}

An independent illustration of this for 2-dimensional data is presented in
Fig.~\ref{fig:comp1}, which shows a field of artificial Gaussian clumps,
with the resulting clump assignment arrays produced by FellWalker and
CLUMPFIND\footnote{In cases such as this, where the clumps are known \emph{a
priori} to be Gaussian in shape, algorithms that assume a Gaussian model
for each clump - such as the GaussClumps algorithm - will provide a better
decomposition of merged sources than either FellWalker or CLUMPFIND. However,
in general the intrinsic shape of each source within a real data array is
unknown. In such cases, assuming a Gaussian shape is likely to require each
source to be split into many overlapping Gaussians in order to match a
real non-Gaussian total source profile.}. The input data image is an array
of $500\times500$ pixels containing 200 circular clumps distributed randomly
across the image\footnote{Except none are allowed to touch an edge of the
image}. Each clump has a Gaussian profile with a Full Width at Half Maximum
of 15 pixels. The clump peak values are distributed uniformly between 30 and
100, and Gaussian noise of standard deviation 15 is added to the image.
Both FellWalker and CLUMPFIND are run with the default parameter values
supplied by CUPID (\emph{FellWalker.MinDip}=3.RMS,
\emph{FellWalker.MaxJump}=4, \emph{ClumpFind.DeltaT}=2.RMS,
\emph{ClumpFind.TLow}=2.RMS).

It can be seen that the two algorithm have different problems; CLUMPFIND
is splitting each real clump into several parts, but FellWalker is
possibly failing to detect some of clumps that are visible by eye. FellWalker
detects 133 clumps\footnote{Since some of the 200 real sources will, by
chance, overlay each other very closely, we cannot expect all 200 clumps
to be detected.} and CLUMPFIND detects 1239.

\begin{figure}
\includegraphics[width=\columnwidth]{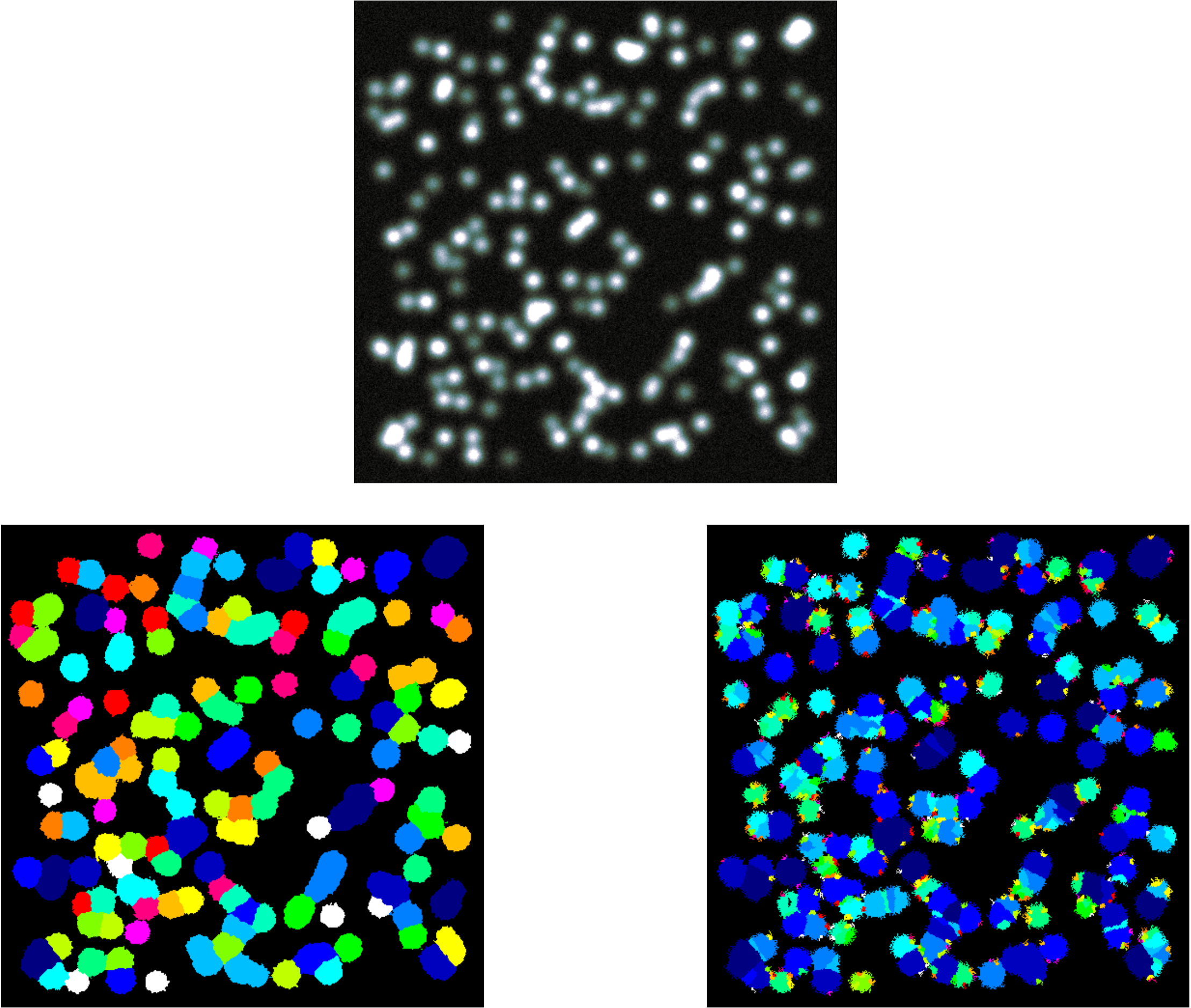}
\caption{The effect of reducing the noise level in the artificial data
by a factor of 5.}
\label{fig:comp2}
\end{figure}

This tendency for CLUMPFIND to split sources seems to be worse for low
signal-to-noise data. If we create a second field of artificial data with
a lower RMS noise level (3 instead of 15), the clump assignment arrays
shown in Fig.~\ref{fig:comp2} are created. CLUMPFIND is still tending to
fragment the edges of clumps into many tiny detections, but the bulk of
the interior of each clump is now left unfragmented.

If either algorithm splits clumps into several parts, not only will it
produce too many clumps, but the total data sum in each clump (\emph{i.e.}
the sum of the pixel values within the clump) will on
average be too low. A useful tool for measuring the performance of these
algorithms is therefore the distribution of the measured total data sum
in each clump compared to the expected distribution, based on knowledge
of the clumps in the artificial data. In order to simplify such a
comparison, all the artificial clumps can be made identical (\emph{i.e.}
have the same peak amplitude and size), so that they all have the same
total data sum. In this case, the distribution of measured total data
sums should be peaked at the expected value, but will always have a tail
of higher-valued clumps due to the random positioning of clumps causing
some clumps to overlay each other. However an optimal clump-finding
algorithm should not produce any significant tail of lower-valued clumps.

The next sections describes the results of many such comparisons,
performed with a range of different signal-to-noise ratios, and with
different clump-finding parameter values.

\subsection{The Artificial Data}
The input data for each test was a two-dimensional image of $500\times500$
pixels containing 200 randomly positioned Gaussian clumps, all with the
same peak value of 100 (arbitrary units) and the same FWHM of 15 pixels
\footnote{Note, this test data is \emph{not} that shown in figures
\ref{fig:comp1} and \ref{fig:comp2} in which the clumps have \emph{different}
peak values.}. Gaussian noise was then added, with a different noise
level for each of eight successive set of tests. The eight noise levels
used were spread evenly between 2 and 16.

\subsection{The Clump-finding Parameters}
For each of the eight different noise levels, the data was analysed
multiple times by both FellWalker and CLUMPFIND, using different values
for the main clump-finding parameters in each case. For CLUMPFIND, the
DeltaT parameter (the gap between contour levels) was varied from twenty
times the noise level down to one times the noise level. For FellWalker,
the MaxJump parameter (the minimum spatial separation between distinct
peaks) was varied between 2 and 14 pixels. In addition, the MinDip
parameter (the smallest dip in height allowed between distinct peaks) was
varied between 0 and 5 times the noise level.

The other main parameter common to both algorithms is the threshold
``sea-level'' below which all pixel values are ignored. This was fixed at
two times the noise level for all tests of both algorithms.

The default implementation of CLUMPFIND provided by the Starlink CUPID
package follows the description of the algorithm contained in
\citet{1994Williams}. The IDL version of CLUMPFIND distributed by Williams
includes some enhancements to the published algorithm. These are also
available in the CUPID version, but are disabled by default. They were,
however, enabled for the purposes of the current comparison.

Full lists of parameters used for these tests are given in
appendix~\ref{app:configs}.

\subsection{Measuring Performance}
In each test, the artificial data consists of a collection of identical
clumps. So the initial expectation is that a reliable clump-finding
algorithm should produce a set of measured clumps each of which has the
same total data sum, and that the total data sum of each measured clump
should be the same as that of an artificial clump. We refer to the ratio
of the measured total data sum to the expected total data sum for a
single clump as the ``gain'' (\emph{i.e.} a measured clump with a gain
greater than unity has a measured total data sum in excess of the expected
total data sum). So for an ideal clump-finder we would expect all gains values
to be unity.

However, this will not be the case in practice for two different reasons
(excluding the random variations caused by the addition of noise to the
data):

\begin{enumerate}

\item Because of the random position of clumps, some clumps will have
spatial overlap to a greater or lesser extent. If the overlap is small,
then a good clump-finding algorithm should be able to resolve them. But
for larger overlaps, and larger noise levels, it becomes progressively
more difficult to resolve overlapping clumps. In the limiting case of
exactly co-incident clumps, it is clearly impossible for any algorithm
to resolve them. For this reason, we expect to see a tail of high-valued
clumps, although it is difficult to quantify the expected size of this
tail.

\item Each algorithm ignores pixel values below a threshold of two times the
noise level. This means that tests performed at higher noise levels will
set the threshold higher and so will detect clumps with smaller total
data sums. See Fig.~\ref{fig:comp5}.

\begin{figure}
\includegraphics[width=\columnwidth]{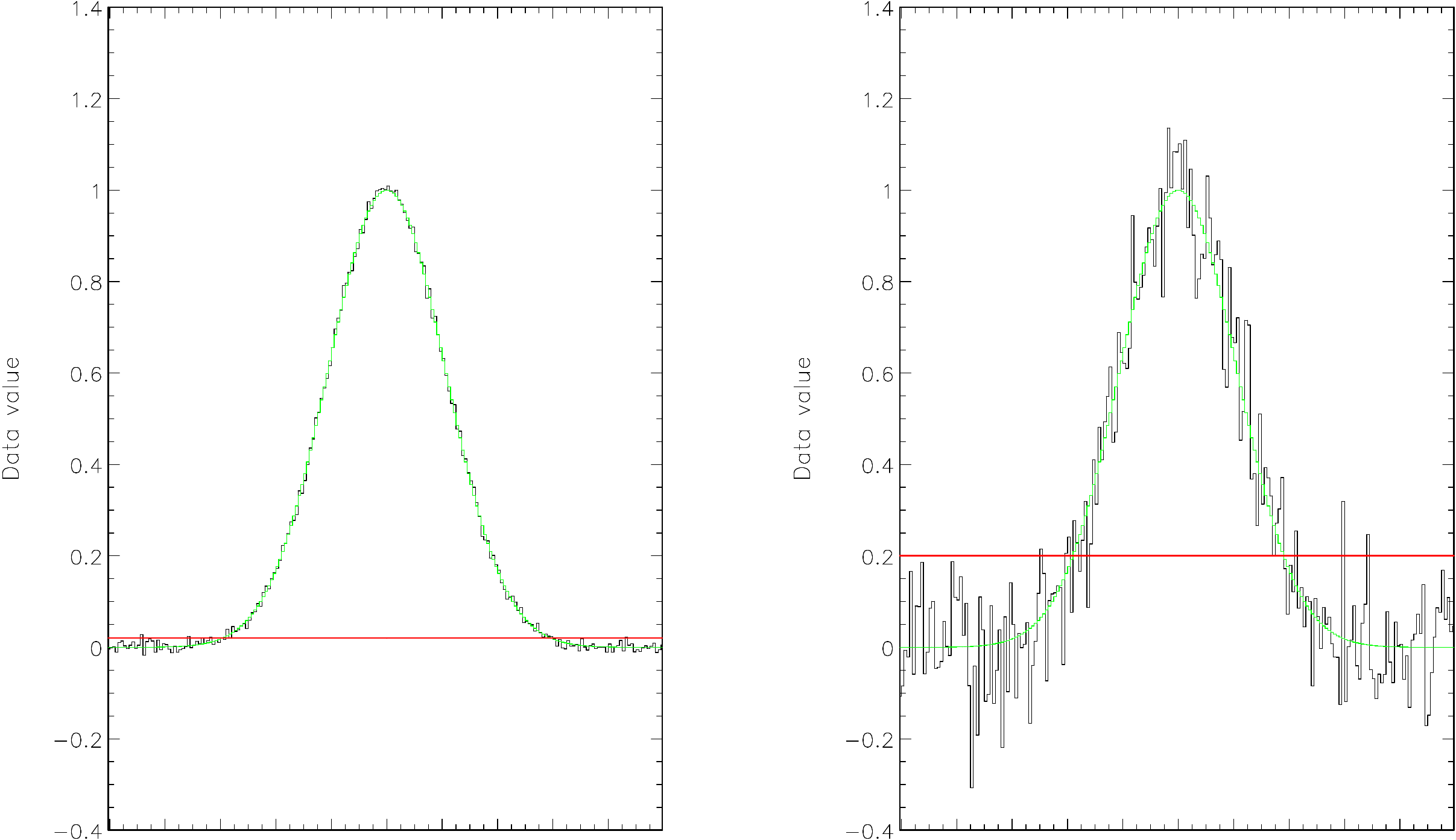}
\caption{Each panel shows the same noise-free Gaussian curve in green.
The black curves are created by adding noise to the Gaussian - 0.02 in
the left panel and 0.1 in the right panel. The red line indicates the two
sigma threshold below which both FellWalker and CLUMPFIND ignore pixels -
0.04 on the left and 0.2 on the right. So the total data sum measured by
both algorithms is the sum of the values above the red line. So it is
expected that they will measure a lower total data sum in the right hand
panel because fewer pixels fall above the red line.
}
\label{fig:comp5}
\end{figure}

\end{enumerate}

Whilst it is difficult to quantify the effects of the first of these two
issues, its impact can be reduced by using the median of the individual
gain values, rather than the mean, as our measure of the performance of each
algorithm. This will reduce the influence of the tail of high-valued clump
data sums.

The second issue can be taken into account by a small re-definition of
the gain value. Instead of defining the gain for a clump as the ratio of
the total data sum measured by FellWalker or CLUMPFIND, to the total data
sum in an artificial Gaussian source, we define it as the ratio of
the measured total data sum, to the total data sum \emph{that falls above the
two sigma threshold} in an artificial Gaussian source. With reference to
Fig.~\ref{fig:comp5}, we are comparing the measured total data sums to
the sum of the green values that fall above the red line.

\subsection{CLUMPFIND Results}
\begin{figure}
\includegraphics[width=\columnwidth]{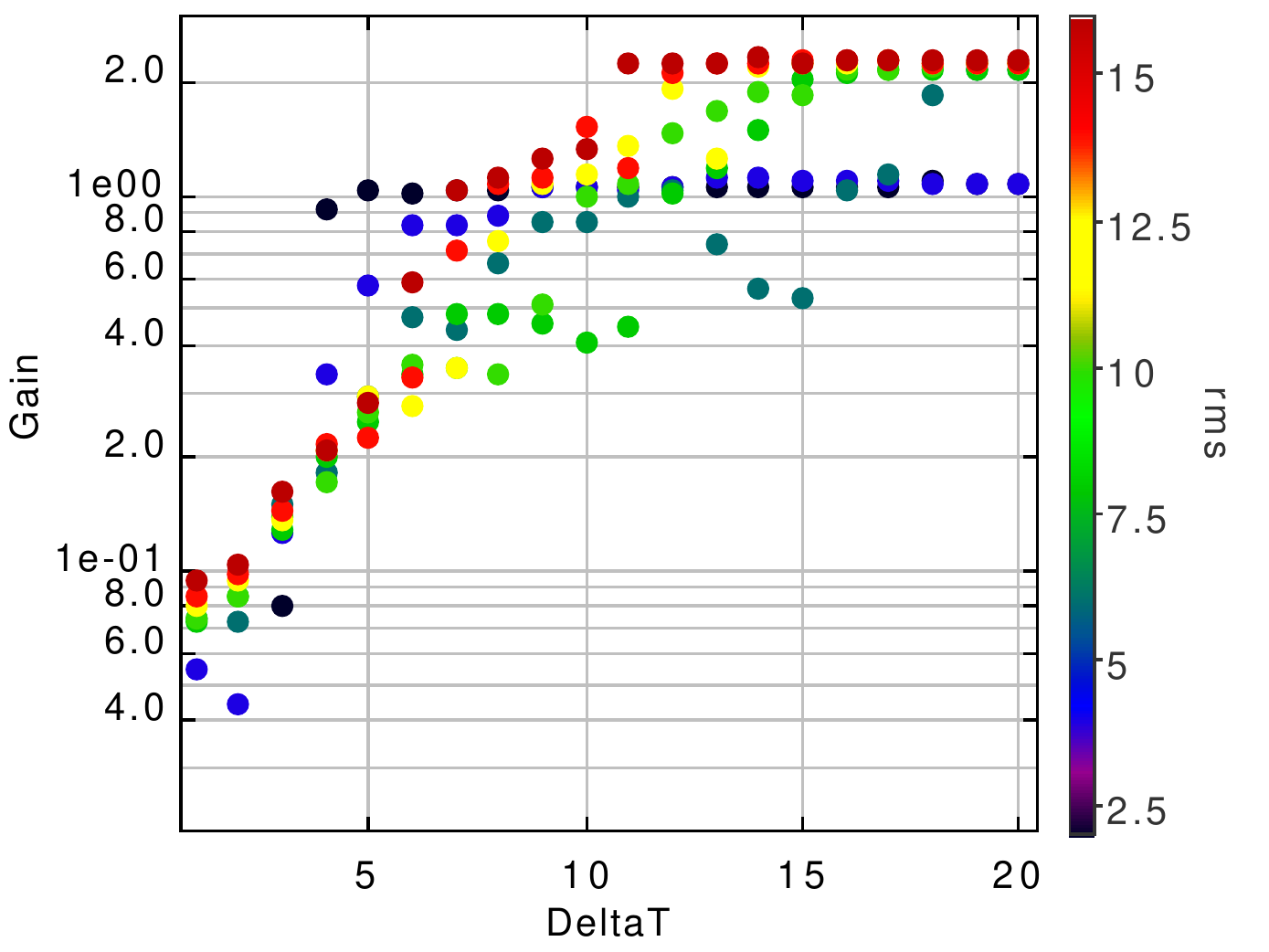}
\caption{The ratio of measured to expected data sum in clumps identified
using CLUMPFIND, as a function of the DelatT parameter value (the gap
between CLUMPFIND contour levels). The DeltaT values are multiples of the
noise level. The colours indicate the noise level, as shown in the colour
bar on the right of the figure.}
\label{fig:cf_results}
\end{figure}

Fig.~\ref{fig:cf_results} shows the ratio of median clump data sum, as
measured by CLUMPFIND, to expected clump data sum at various values of
the DeltaT parameter and for various noise levels. Ideally , we would
hope for a gain of unity (\emph{i.e.} the median measured clump data sum
equalling the expected clump data sum). It can be seen that such a
condition is typically achieved at a DeltaT value around 10 times the
noise level, but that the relationship between DeltaT, Gain and RMS is
quite unpredictable. This supports the findings of \citet{2009Pineda},
who found that CLUMPFIND results were very sensitive to the value of DeltaT.

At the DeltaT value of 2.0 recommended by \citet{1994Williams}, all noise
levels produce clumps which are well under the expected total data sum,
supporting the finding of \citet{2010Watson} that CLUMPFIND tends to
fragment clumps.

\subsection{FellWalker Results}
\begin{figure}
\centering
\subfloat[][]{\includegraphics[trim=0 0 25mm 0,clip,height=0.5\columnwidth]{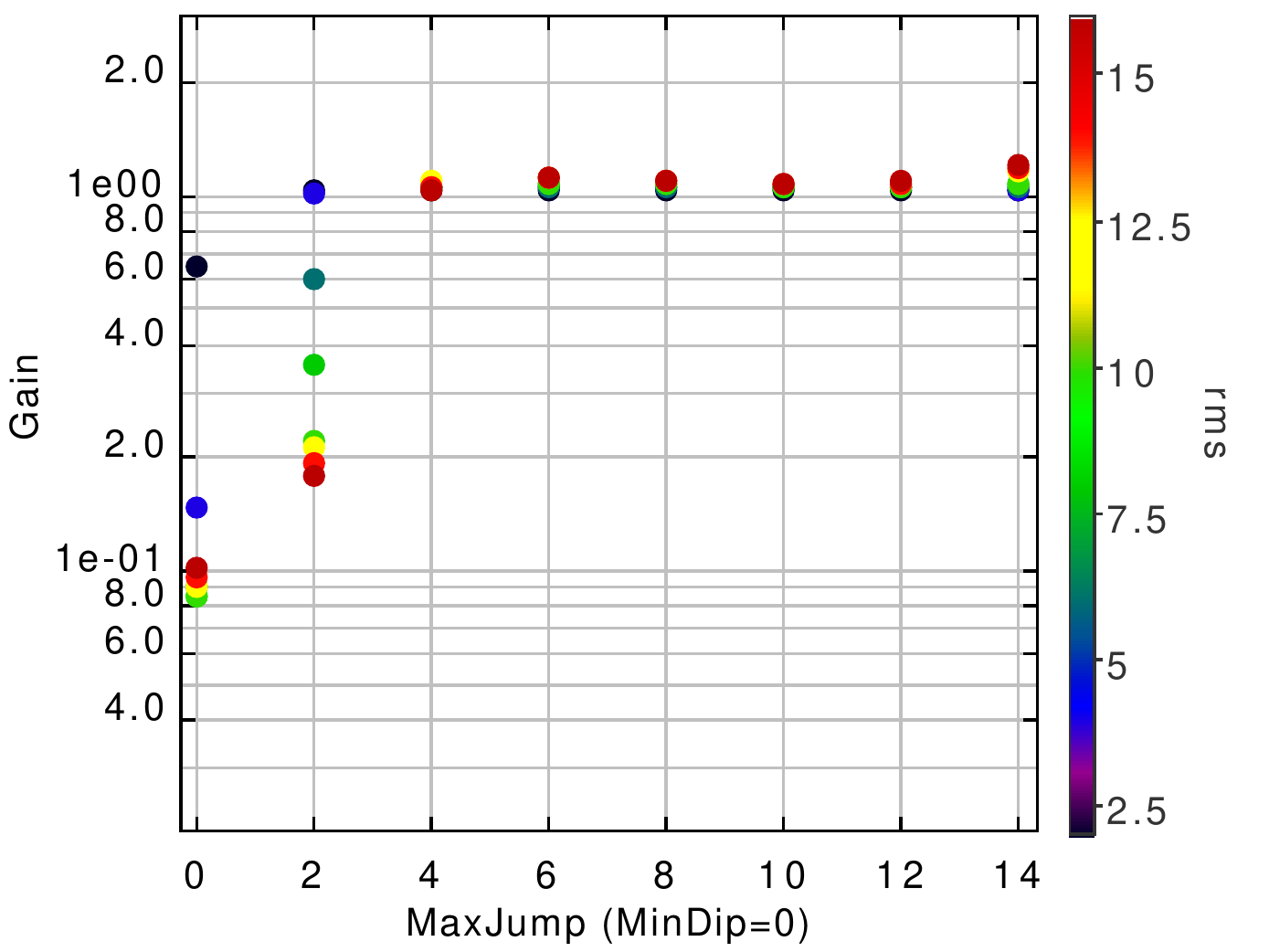}}
\hspace*{10pt}
\subfloat[][]{\includegraphics[trim=20mm 0 25mm 0,clip,height=0.5\columnwidth]{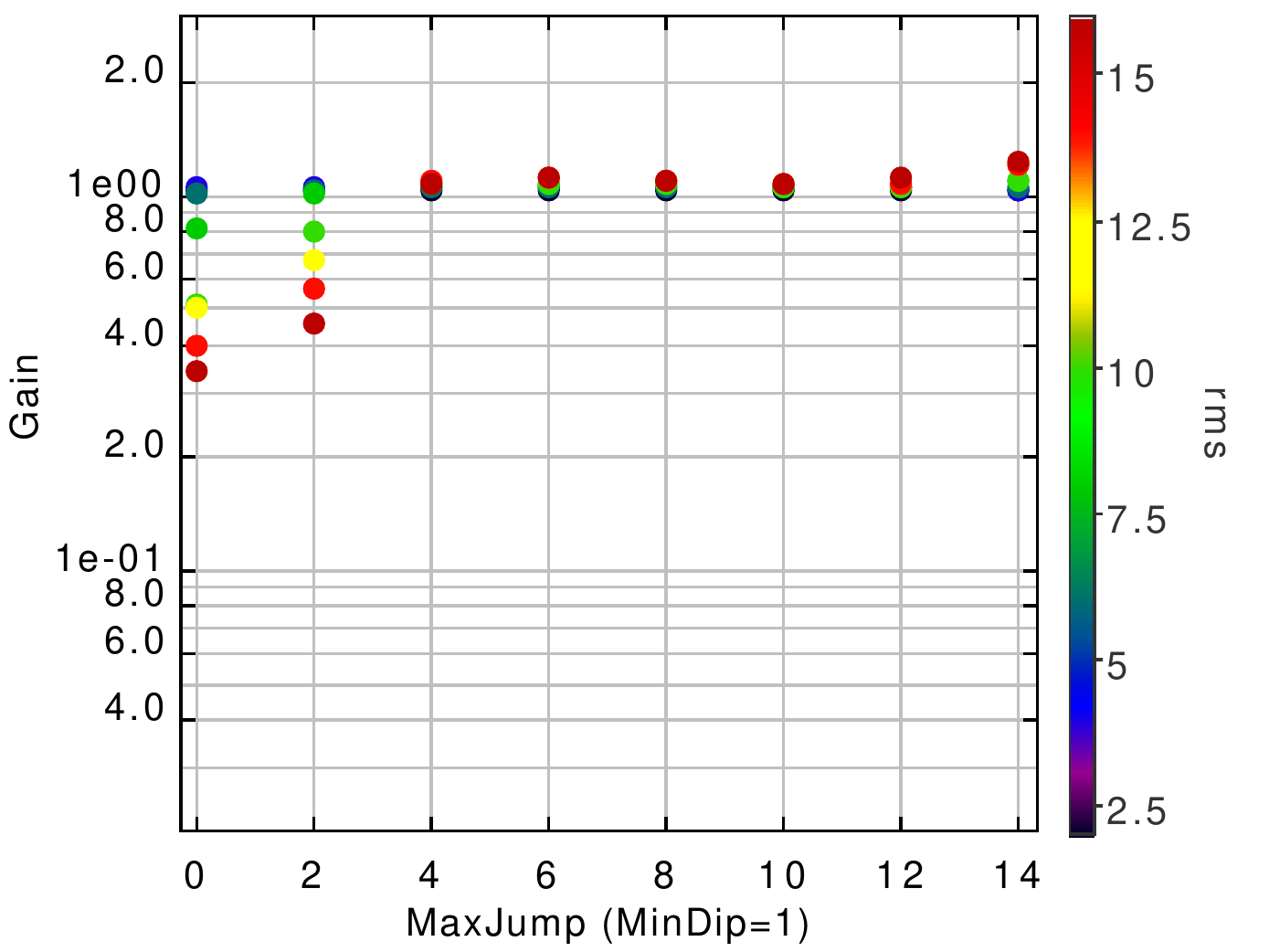}}\\
\subfloat[][]{\includegraphics[trim=0 0 25mm 0,clip,height=0.5\columnwidth]{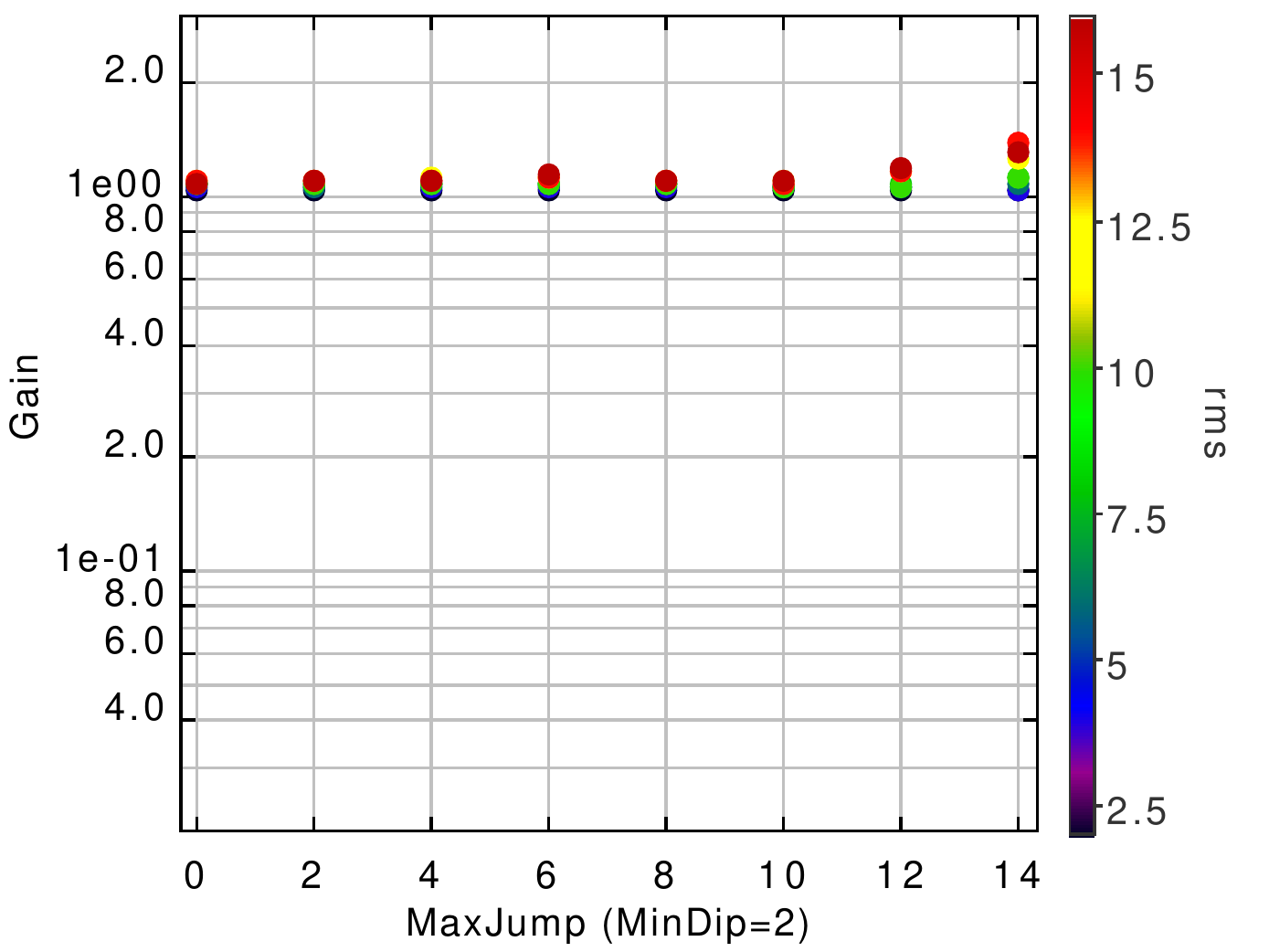}}
\hspace*{10pt}
\subfloat[][]{\includegraphics[trim=20mm 0 25mm 0,clip,height=0.5\columnwidth]{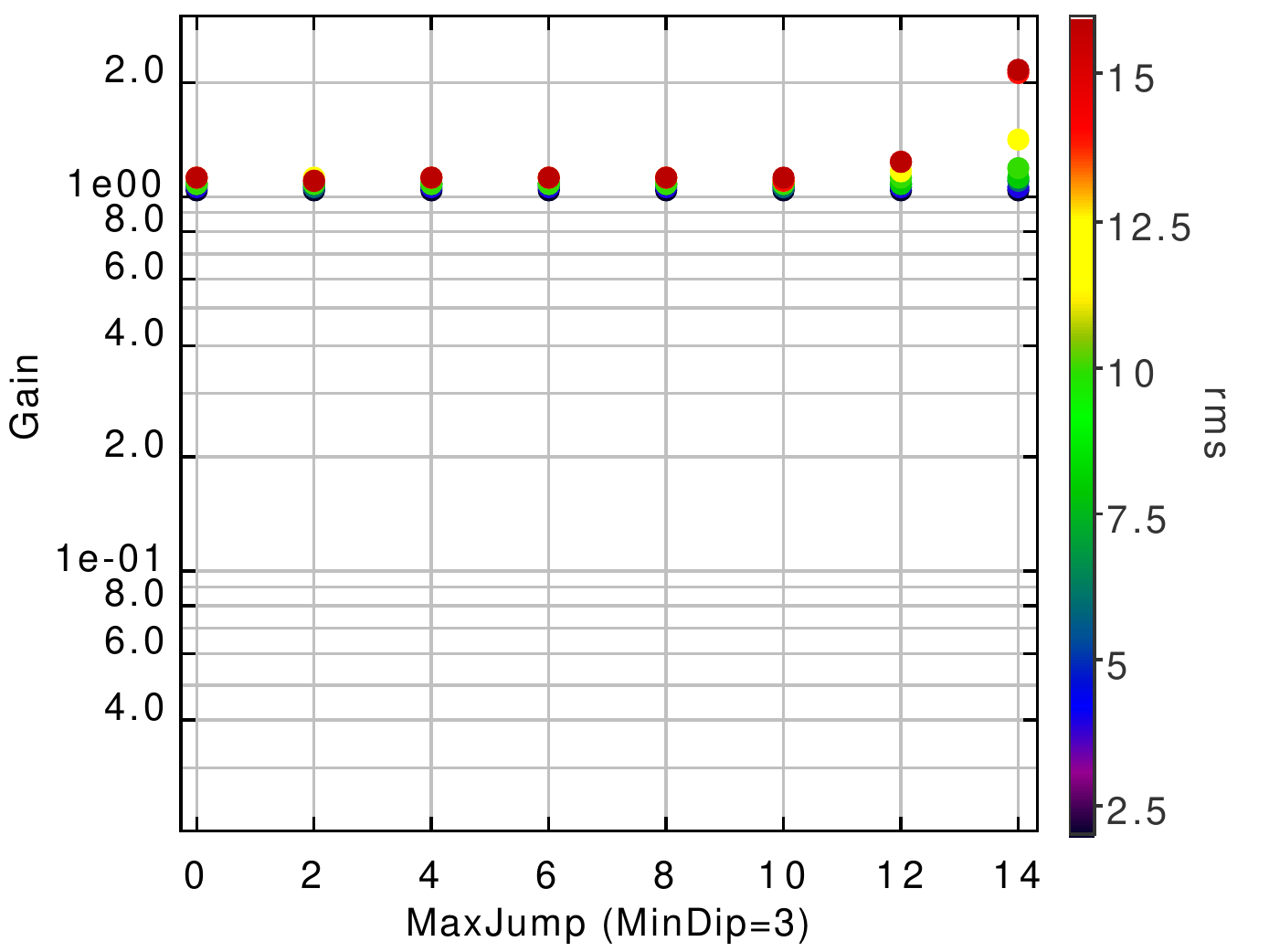}}\\
\subfloat[][]{\includegraphics[trim=0 0 25mm 0,clip,height=0.5\columnwidth]{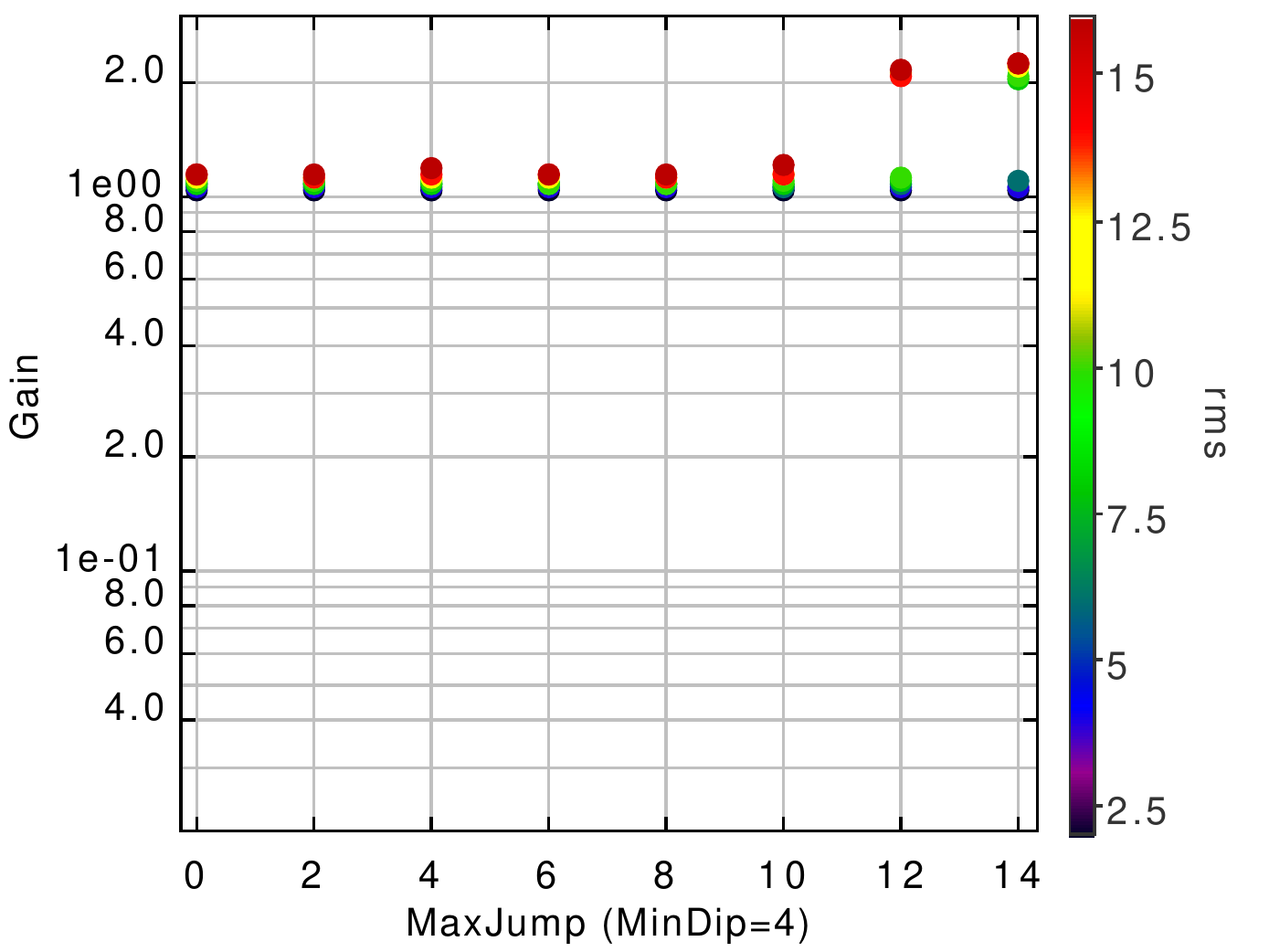}}
\hspace*{10pt}
\subfloat[][]{\includegraphics[trim=20mm 0 25mm 0,clip,height=0.5\columnwidth]{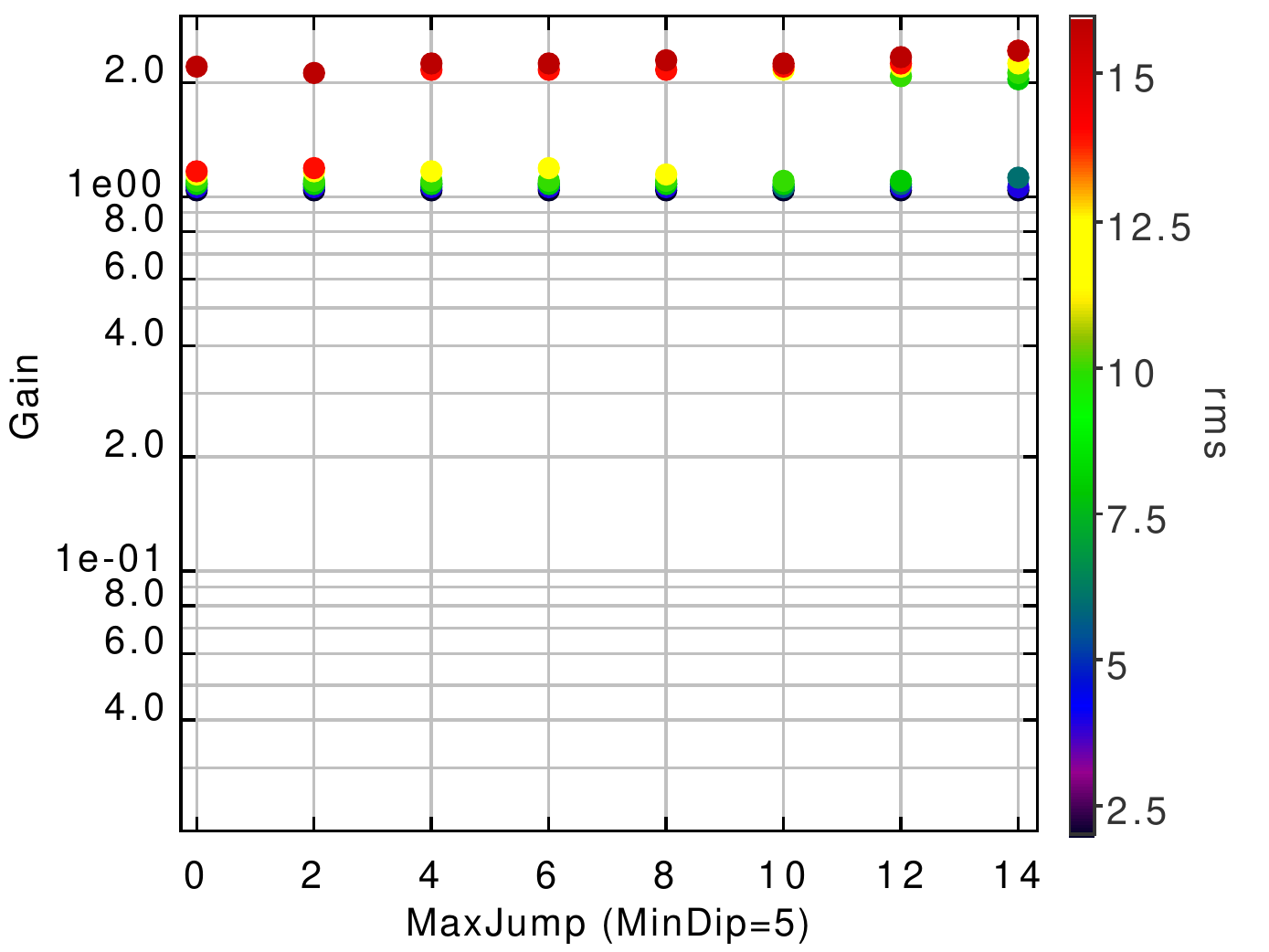}}
\caption{The ratio of measured to expected data sum in clumps identified
using FellWalker, as a function of the MaxJump parameter value (the
minimum spatial distance between distinct peaks, in pixels), and the
MinDip parameter value (the minimum dip between distinct peaks, as a
multiple of the noise level). The colours indicate the noise level, as
shown in Fig.~\ref{fig:cf_results}.}
\label{fig:fw_results}
\end{figure}

Fig.~\ref{fig:fw_results} shows the ratio of median clump data sum, as
measured by FellWalker, to expected clump data sum at various values of
the MaxJump and MinDip parameters and for the same noise levels shown in
Fig.~\ref{fig:cf_results}. It can be seen that in these tests, FellWalker
produces clumps with data sums that are much more consistent and closer
to the expected data sum than CLUMPFIND.

It can be seen that at small MaxJump and small MinDip, FellWalker splits
clumps into small fragments. The small MaxJump value causes noise spikes
to be interpreted as peaks, and the small MinDip value then prevents these
``peaks'' from being merged. At the other extreme, large MaxJump values
allow the walk process to jump between real peaks, thus merging them
together into clumps that are larger than expected. This effect is worse
at larger MinDip values.

But between these extremes, values of MaxJump between 4 and 10, and MinDip
between 0 and 4 times the noise level produce predictable gain values that
are close to the expected value of unity, meaning that the median clump data
sum is close to the value expected on the basis of the known properties
of the artificial clumps.

\subsection{Comparisons Using Three-Dimensional Data}

\begin{figure}
\includegraphics[width=\columnwidth]{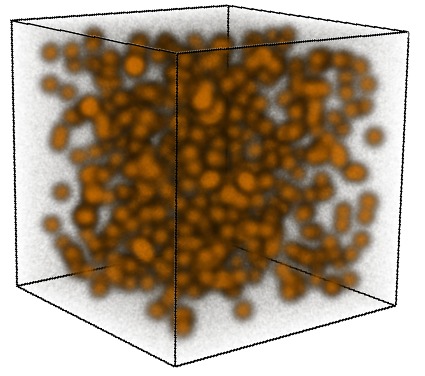}
\caption{A volume-rendering of a typical artificial data cube used to
compare the performance of CLUMPFIND and FellWalker on 3-dimensional data.}
\label{fig:true_6}
\end{figure}

The above comparison of FellWalker and CLUMPFIND was done using
2-dimensional data. This section presents results of a similar test using
3-dimensional data. Each test data cube has dimensions of $200\times200\times200$
pixels, and contains 600 artificial Gaussian clumps each with a width
(FWHM) of 10 pixels and identical peak value of 100. In all other
respects, the 3-dimensional tests were done in exactly the same way and
with the same parameter ranges as the 2-dimensional tests.
Fig.~\ref{fig:true_6} shows a volume rendering of the artificial data
cube with noise level of 6.0

\begin{figure}
\includegraphics[width=\columnwidth]{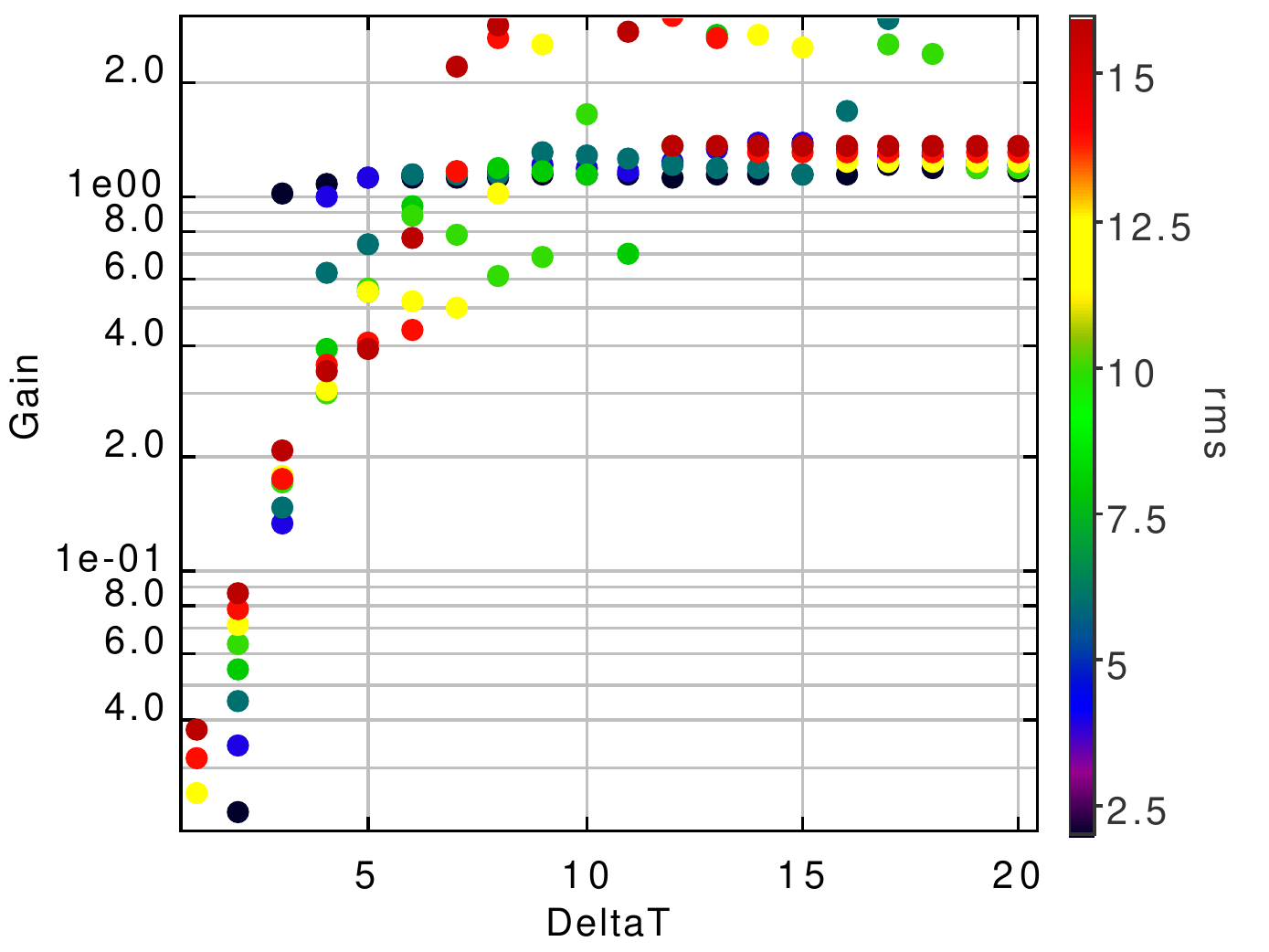}
\caption{The ratio of measured to expected data sum in 3-dimensional clumps
identified using CLUMPFIND, as a function of the DelatT parameter value.
The DeltaT values are multiples of the noise level. The colours indicate
the noise level, as shown in the colour
bar on the right of the figure.}
\label{fig:cf_results_3d}
\end{figure}

Fig.~\ref{fig:cf_results_3d} again shows the ratio of median clump data
sum, as measured by CLUMPFIND, to expected clump data sum at the same set
of DeltaT and noise levels as used in the 2-dimensional tests.

\begin{figure}
\centering
\subfloat[][]{\includegraphics[trim=0 0 25mm 0,clip,height=0.5\columnwidth]{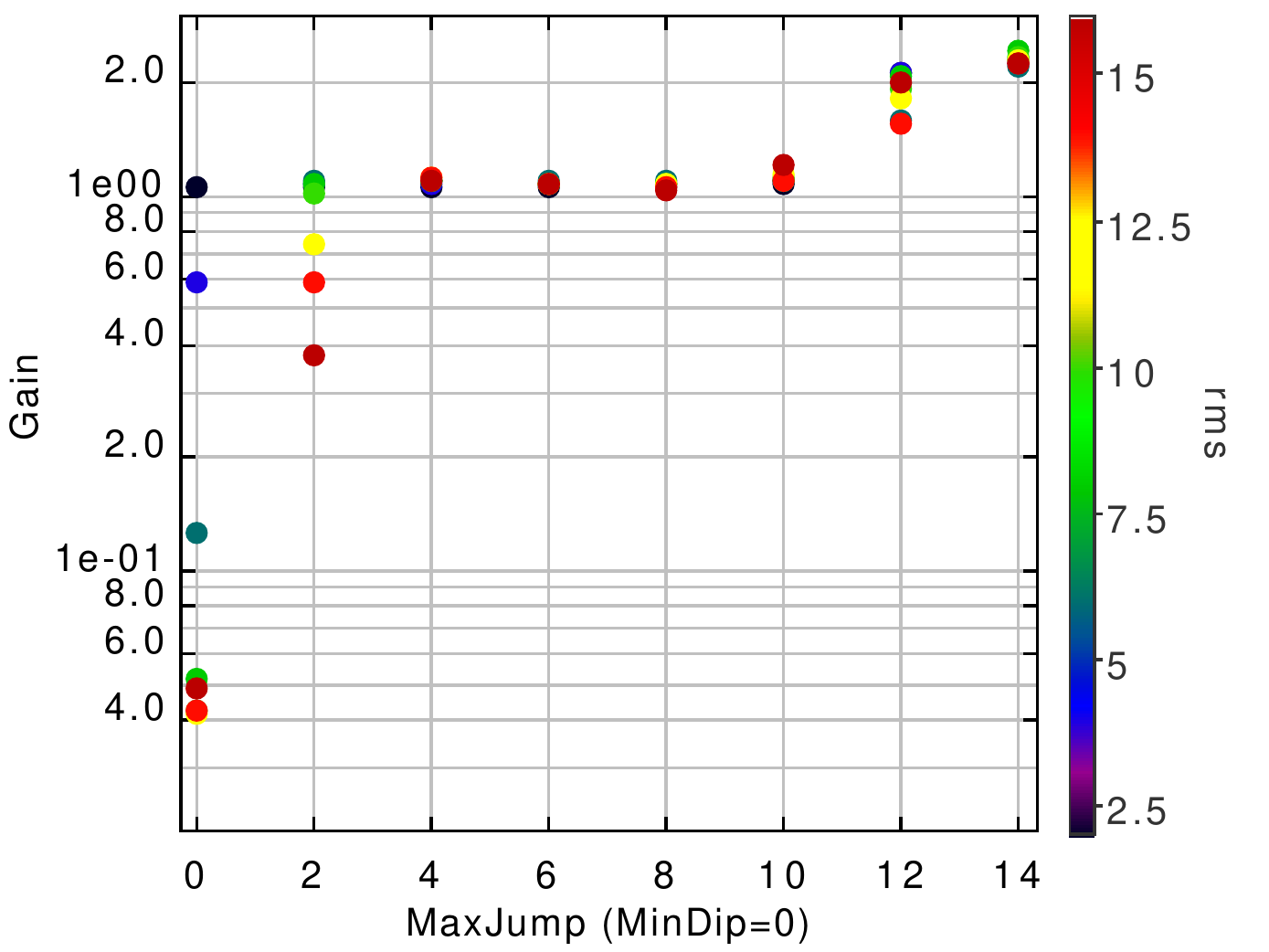}}
\hspace*{10pt}
\subfloat[][]{\includegraphics[trim=20mm 0 25mm 0,clip,height=0.5\columnwidth]{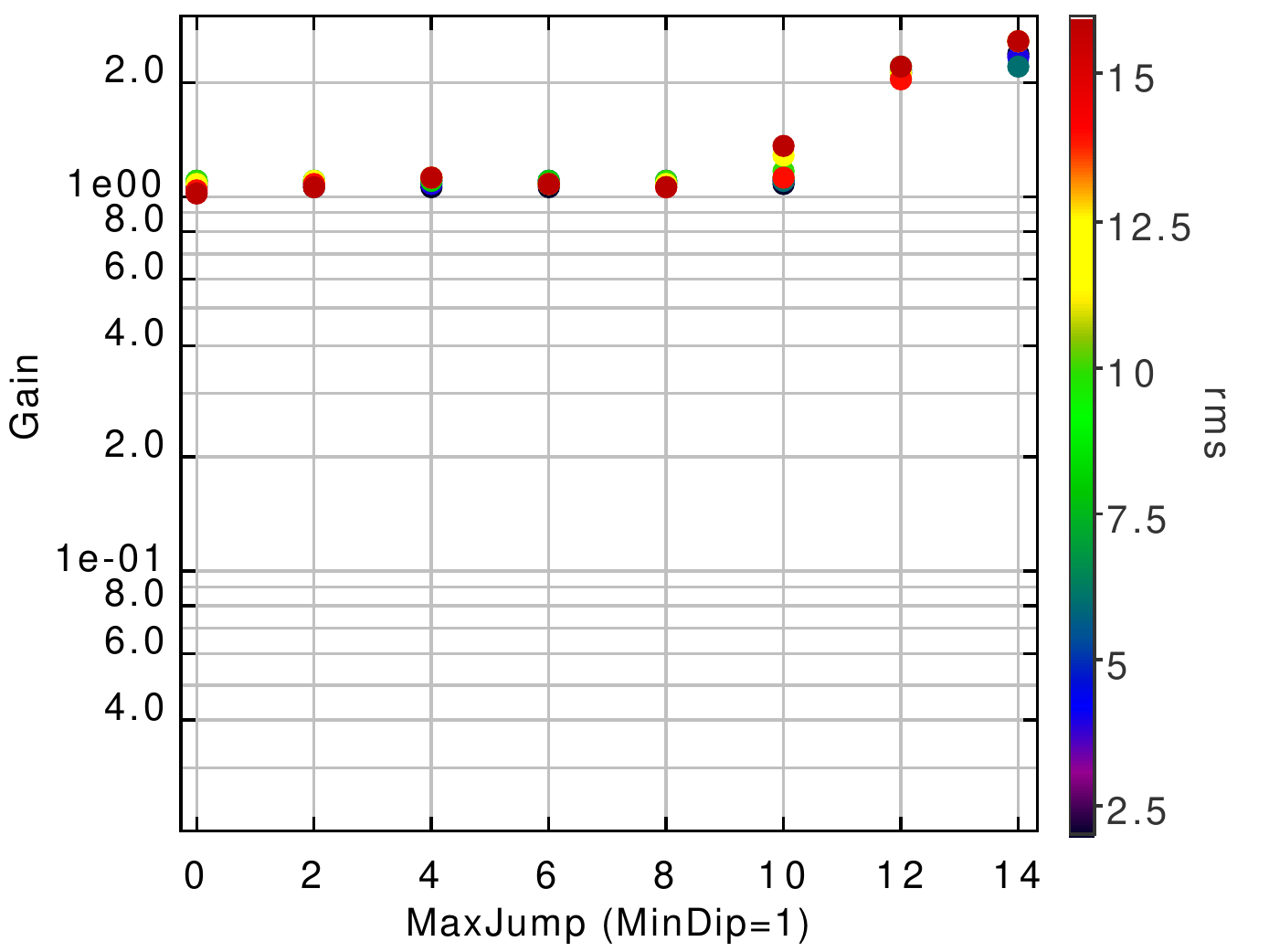}}\\
\subfloat[][]{\includegraphics[trim=0 0 25mm 0,clip,height=0.5\columnwidth]{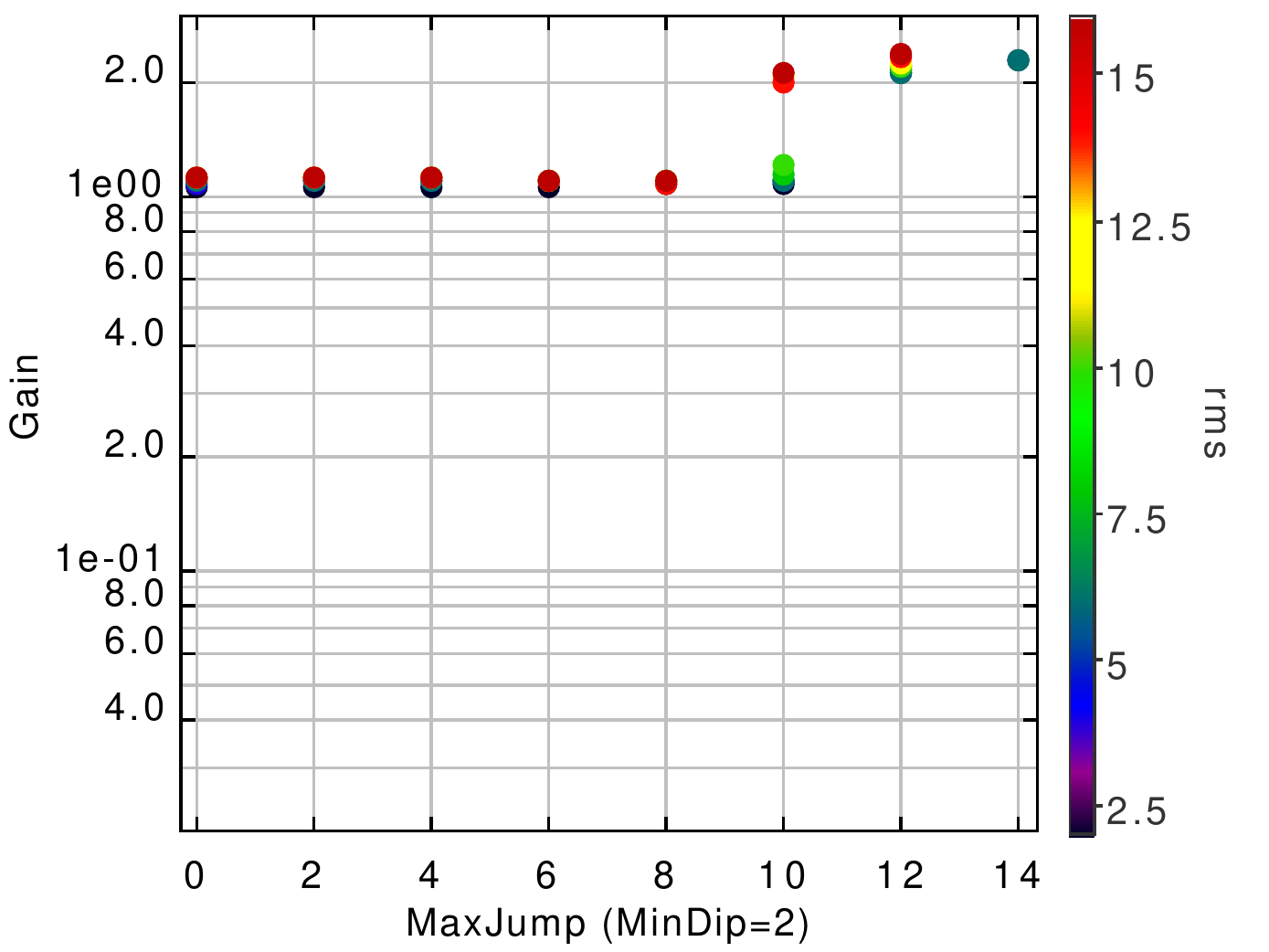}}
\hspace*{10pt}
\subfloat[][]{\includegraphics[trim=20mm 0 25mm 0,clip,height=0.5\columnwidth]{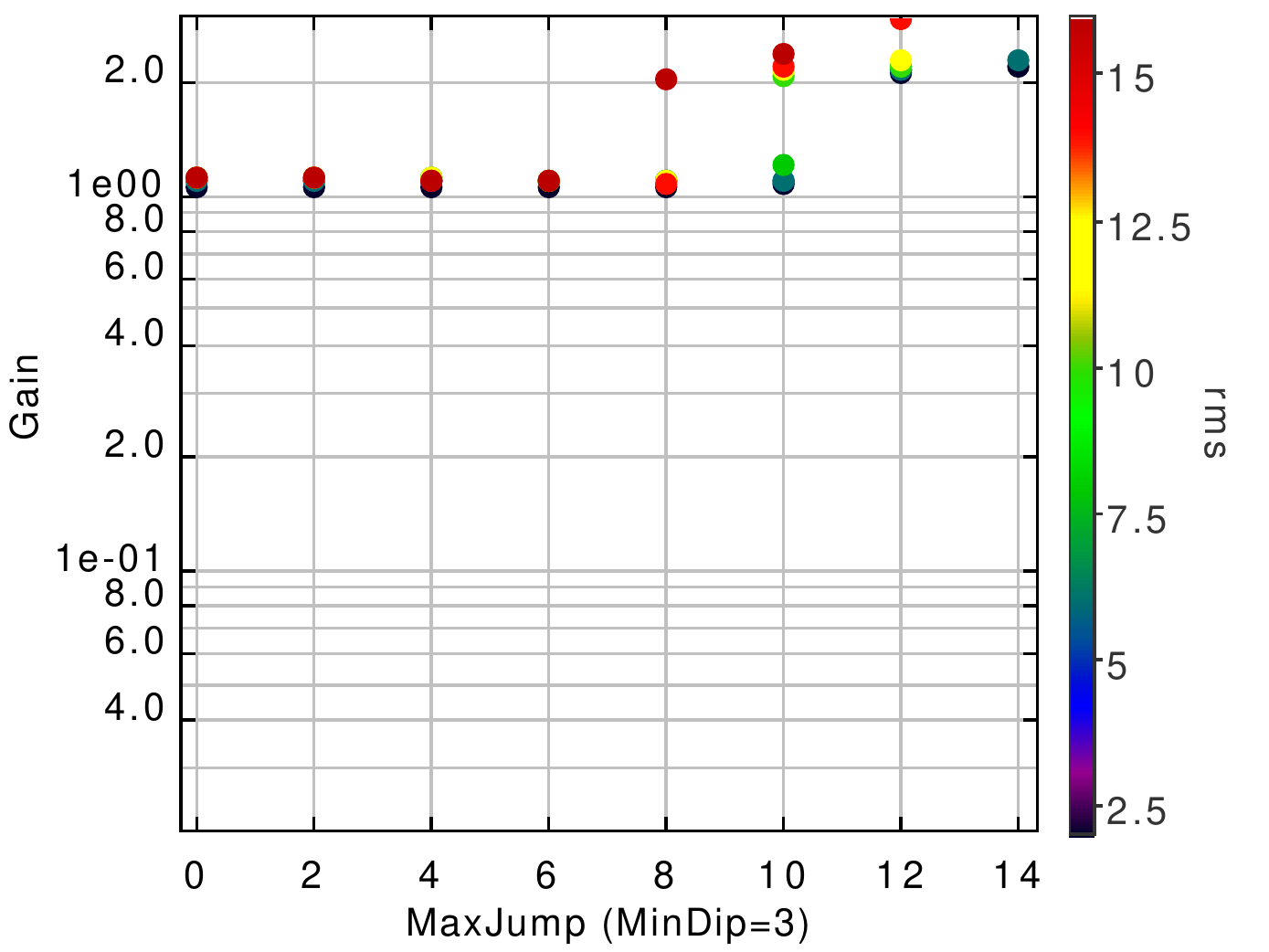}}\\
\subfloat[][]{\includegraphics[trim=0 0 25mm 0,clip,height=0.5\columnwidth]{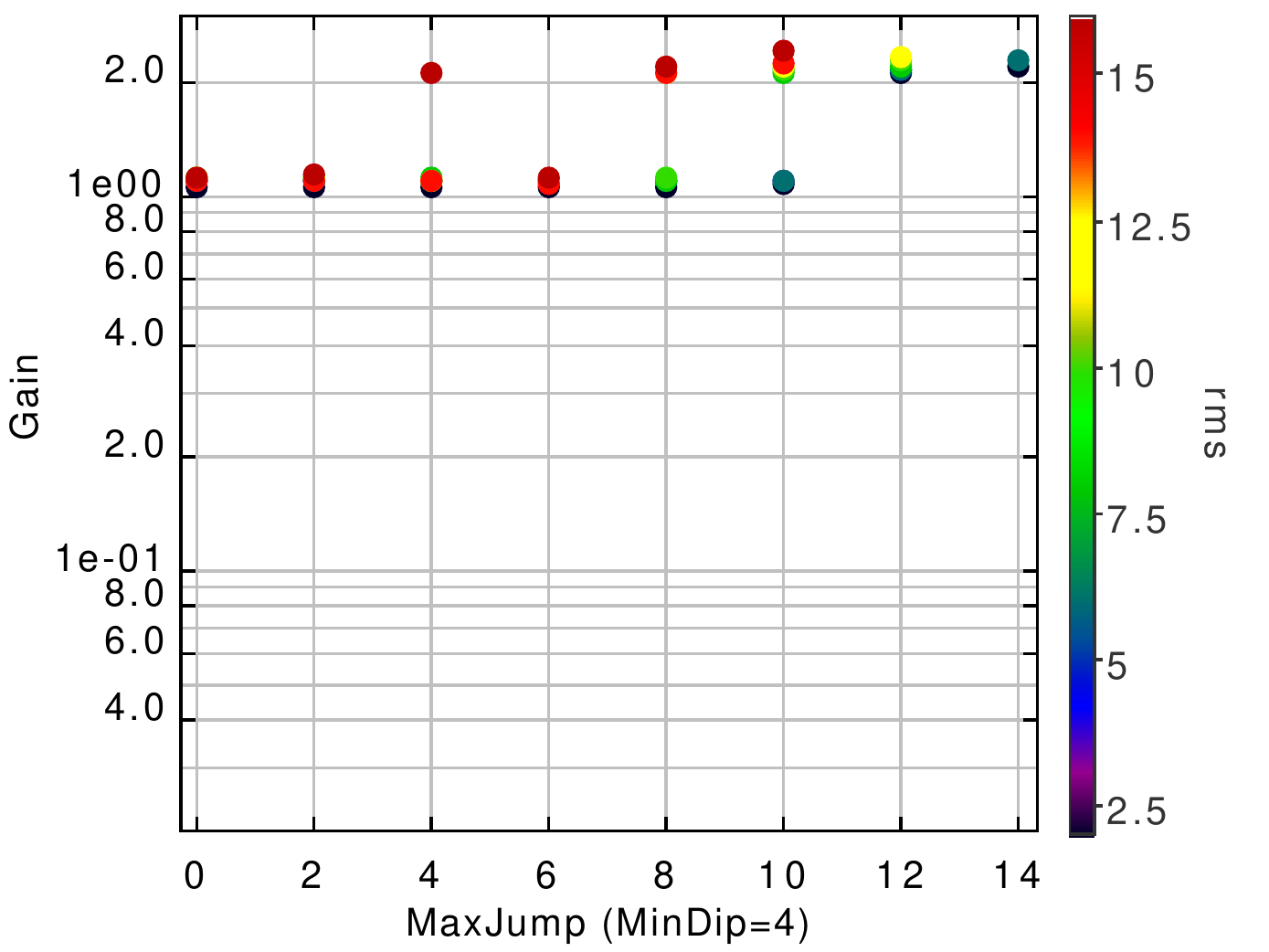}}
\hspace*{10pt}
\subfloat[][]{\includegraphics[trim=20mm 0 25mm 0,clip,height=0.5\columnwidth]{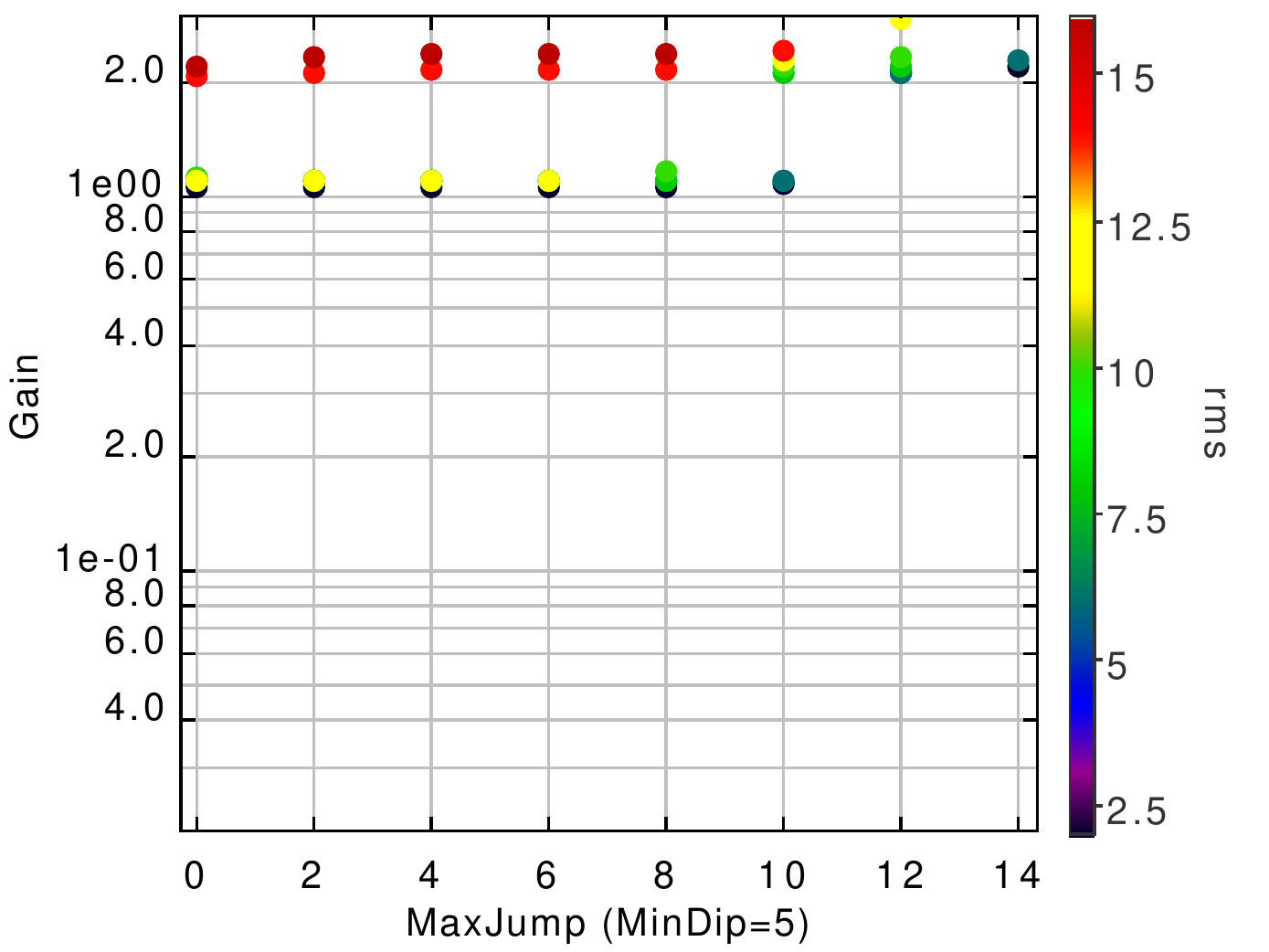}}
\caption{The ratio of measured to expected data sum in 3-dimensional clumps identified
using FellWalker, as a function of the MaxJump and MinDip parameter value.
The colours indicate the noise level, as shown in
Fig.~\ref{fig:cf_results_3d}.}
\label{fig:fw_results_3d}
\end{figure}

Fig.~\ref{fig:fw_results_3d} shows the ratio of median clump data sum, as
measured by FellWalker, to expected clump data sum at the same set of
MaxJump values, MinDip values and noise levels used for the 2-dimensional
tests.

It can be seen that these plots are very similar to the earlier plots
showing the corresponding results for 2-dimensional data. Again,
CLUMPFIND results are very sensitive to DeltaT and are quite
unpredictable, with the recommended DeltaT value (2$*$RMS) splitting
clumps into pieces with smaller than expected data sums. FellWalker is
less sensitive to parameter settings and is more reliable at reproducing
the expected clump data sum.

The main difference is that FellWalker results are more sensitive to the
value of MaxJump in the 3-dimensional case. The best results are obtained
for MinDip between 1.0 and 3.0 and MaxJump less than 8. The default
values supplied by CUPID are 2.0 for MinDip and 4 for MaxJump.

\subsection{Execution Times}
For both 2-dimensional and 3-dimensional data, FellWalker is between 1 and
2 orders of magnitude faster than the CUPID implementation of CLUMPFIND,
which is itself about an order of magnitude faster than the IDL
implementation of CLUMPFIND. Currently, CUPID uses single-threaded
implementations of all clump-finding algorithms.

\section{Future Enhancements to FellWalker}
Future work on the FellWalker algorithm is planned to address two small
problems with the current implementation:

\begin{itemize}
\item The cleaning process described in section~\ref{sec:cleaning} can
sometimes cause clumps to be split into two or more dis-contiguous parts.
This can occur for ``dog-bone'' shaped clumps that have a narrow
steep-sided central ridge that widens out at the two ends. In such cases,
the cleaning process can sometimes erode pixels from the central ridge
causing a gap to appear between the two wider end regions.

\item As described in section~\ref{sec:raw}, when a walk reaches a local
maximum, an attempt is made to find a higher pixel value within a small
box centred on the local maximum, and if found, the walk continues from this
higher pixel. However, no check is made that the higher pixel is on the
same ``island''\footnote{i.e. a contiguous group of
pixels that are all higher than the threshold level.} as the local
maximum. Thus it is possible that walks could jump from one island to
another, and so merge clumps together that are in fact distinct.
\end{itemize}

\section{Acknowledgements}

I would like to thank Tim Jenness and Frossie Economou for encouraging
and promoting the development of the CUPID package and the FellWalker
algorithm, and for providing helpful comments on this paper.

The Starlink software is currently maintained by the Joint Astronomy
Centre, Hawaii with support from the UK Science and Technology
Facilities Council.

\appendix

\section{Configuration Parameter Values}
\label{app:configs}
The following two sections shows the CUPID commands that were
used to perform the 2-dimensional tests described in section
section~\ref{sec:compare}. Also shown are the contents of the text files
that specify the values used for the additional configuration parameters
required by CUPID. See the CUPID documentation \citep{SUN255} for
descriptions of these commands and parameters. In both cases,
\texttt{data.sdf} is the data file containing the artificial clumps,
\texttt{mask.sdf} is the generated clump pixel mask and \texttt{cat.fit}
is the generate clump catalogue. For each specific test, different
numerical values were assigned to the variables \texttt{\$rms},
\texttt{\$maxjump}, \texttt{\$mindip} and \texttt{\$deltat}, as described
in section~\ref{sec:compare}.
\subsection{FellWalker}
{\small
\begin{verbatim}
  % findclumps in=data.sdf deconv=no method=fellwalker \
               out=mask.sdf outcat=cat.fit rms=$rms \
               config="'^fw.conf,MaxJump=$maxjump, \
                        MinDip=$mindip*RMS'" \
               wcspar=no backoff=yes perspectrum=no

  % cat fw.conf
     FellWalker.AllowEdge=1
     FellWalker.CleanIter=1
     FellWalker.FlatSlope=1*RMS
     FellWalker.FwhmBeam=0
     FellWalker.MaxBad=0.05
     FellWalker.MinHeight=2*RMS
     FellWalker.MinPix=16
     FellWalker.Noise=2*RMS
     FellWalker.VeloRes=0
\end{verbatim}
}
\subsection{ClumpFind}
{\small
\begin{verbatim}
  % findclumps in=data.sdf deconv=no method=clumpfind \
               out=mask.sdf outcat=cat.fit rms=$rms \
               config="'^cf.conf,DeltaT=$deltat*RMS'" \
               wcspar=no backoff=yes perspectrum=no

  % cat cf.conf
     ClumpFind.Allowedge=1
     ClumpFind.Fwhmbeam=0
     ClumpFind.Idlalg=1
     ClumpFind.Maxbad=1.0
     ClumpFind.Minpix=16
     ClumpFind.Naxis=2
     ClumpFind.Tlow=2*RMS
     ClumpFind.VeloRes=0
\end{verbatim}
}
For the 3-dimensional tests, the value used for \texttt{ClumpFind.Naxis}
was changed from 2 to 3.

\bibliographystyle{model2-names-astronomy}
\bibliography{fellwalker}







\end{document}